\documentclass[twocolumn,twocolappendix]{aastex631}

\usepackage{upgreek}
\usepackage{multirow}

\usepackage{soul}

\begin{document}

\title{Follow-Up Exploration of the TWA 7 Planet-Disk System with JWST NIRCam}

\correspondingauthor{Katie A. Crotts}
\email{kcrotts@stsci.edu}

\author[0000-0003-4909-256X]{Katie A. Crotts}
\affiliation{Space Telescope Science Institute (STScI), 3700 San Martin Drive, Baltimore, MD 21218, USA}

\author[0000-0001-5365-4815]{Aarynn L. Carter}
\affiliation{Space Telescope Science Institute (STScI), 3700 San Martin Drive, Baltimore, MD 21218, USA}

\author[0000-0002-6964-8732]{Kellen Lawson}
\affiliation{NASA-Goddard Space Flight Center, 8800 Greenbelt Rd, Greenbelt, MD 20771, USA}

\author[0000-0001-5864-9599]{James Mang}
\altaffiliation{NSF Graduate Research Fellow}
\affiliation{Department of Astronomy, University of Texas at Austin, Austin, TX 78712, USA}

\author[0000-0003-4614-7035]{Beth Biller}
\affiliation{Scottish Universities Physics Alliance, Institute for Astronomy, University of Edinburgh, Blackford Hill, Edinburgh EH9 3HJ, UK} \affiliation{Centre for Exoplanet Science, University of Edinburgh, Edinburgh EH9 3HJ, UK}

\author[0000-0001-8568-6336]{Mark Booth}
\affiliation{UK Astronomy Technology Centre, Royal Observatory Edinburgh, Blackford Hill, Edinburgh EH9 3HJ, UK}

\author[0000-0002-2683-2396]{Rodrigo Ferrer-Chavez}
\affiliation{Center for Interdisciplinary Exploration and Research in Astrophysics (CIERA) and Department of Physics and Astronomy, Northwestern University, Evanston, IL, 60208, USA}

\author[0000-0001-8627-0404]{Julien H. Girard}
\affiliation{Space Telescope Science Institute (STScI), 3700 San Martin Drive, Baltimore, MD 21218, USA}

\author[0000-0002-2189-2365]{Anne-Marie Lagrange}
\affiliation{LIRA, CNRS, Observatoire de Paris, CNRS, Universit\'{e} PSL, 5 Place Jules Janssen, 92190 Meudon, France}
\affiliation{Univ. Grenoble Alpes, CNRS, IPAG, F-38000 Grenoble, France}

\author[0000-0003-2232-7664]{Michael C. Liu}
\affiliation{Institute for Astronomy, University of Hawaii, 2680 Woodlawn Drive, Honolulu HI 96822}

\author[0000-0002-5352-2924]{Sebastian Marino}
\affiliation{Department of Physics and Astronomy, University of Exeter, Stocker Road, Exeter EX4 4QL, UK}

\author[0000-0001-6205-9233]{Maxwell A. Millar-Blanchaer}
\affiliation{Department of Physics, University of California, Santa Barbara, CA 93106, USA}

\author[0000-0001-6098-3924]{Andy Skemer}
\affiliation{Department of Astronomy \& Astrophysics, University of California, Santa Cruz, CA 95064, USA }

\author[0000-0002-1652-420X]{Giovanni M. Strampelli}
\affiliation{Space Telescope Science Institute (STScI), 3700 San Martin Drive, Baltimore, MD 21218, USA}

\author[0000-0003-0774-6502]{Jason Wang}
\affiliation{Center for Interdisciplinary Exploration and Research in Astrophysics (CIERA) and Department of Physics and Astronomy, Northwestern University, Evanston, IL, 60208, USA}

\author[0000-0002-4006-6237]{Olivier Absil}
\affiliation{STAR Institute, Universit\'e de Li\`ege, 19c All\'ee du Six Ao\^ut, 4000 Li\`ege, Belgium}

\author[0000-0001-6396-8439]{William O. Balmer}
\affiliation{Department of Physics \& Astronomy, Johns Hopkins University, 3400 N. Charles Street, Baltimore, MD 21218, USA}
\affiliation{Space Telescope Science Institute (STScI), 3700 San Martin Drive, Baltimore, MD 21218, USA}

\author[0009-0000-0303-2145]{Rapha\"el Bendahan-West}
\affiliation{Department of Physics and Astronomy, University of Exeter, Stocker Road, Exeter EX4 4QL, UK}

\author[0000-0002-7325-5990]{Ellis Bogat}
\affiliation{Department of Astronomy, University of Maryland, College Park, 4296 Stadium Dr, College Park, MD 20742, USA}

\author[0000-0001-5831-9530]{Rachel Bowens-Rubin}
\affiliation{Department of Astronomy, University of Michigan, Ann Arbor, MI 48109, USA}
\affiliation{Eureka Scientific Inc., 2542 Delmar Ave., Suite 100, Oakland, CA 94602, USA}

\author[0000-0003-4022-8598]{Ga\"el Chauvin}
\affiliation{Max-Planck-Institut für Astronomie, K\"onigstuhl 17, 69117 Heidelberg, Germany}
\affiliation{Laboratoire Lagrange, Universit\'e C\^ote d'Azur, CNRS, Observatoire de la C\^ote d'Azur, 06304 Nice, France}

\author[0000-0002-2428-9932]{Cl\'emence Fontanive}
\affiliation{Trottier Institute for Research on Exoplanets, Universit\'e de Montr\'eal, Montr\'eal H3C 3J7, Qu\'ebec, Canada}

\author[0000-0003-4557-414X]{Kyle Franson}
\altaffiliation{NSF Graduate Research Fellow}
\affiliation{Department of Astronomy, University of Texas at Austin, Austin, TX 78712, USA}

\author[0000-0003-2769-0438]{Jens Kammerer}
\affiliation{European Southern Observatory, Karl-Schwarzschild-Str. 2, 85748 Garching, Germany}

\author[0000-0002-0834-6140]{Jarron Leisenring}
\affiliation{Department of Astronomy / Steward Observatory, University of Arizona, Tucson, AZ 85721, USA}

\author[0000-0002-4404-0456]{Caroline V. Morley}
\affiliation{Department of Astronomy, University of Texas at Austin, Austin, TX 78712, USA}

\author[0000-0002-4388-6417]{Isabel Rebollido}
\affiliation{European Space Agency (ESA), European Space Astronomy Centre (ESAC), Camino Bajo del Castillo s/n, 28692 Villanueva de la Ca\~nada, Madrid, Spain}

\author[0000-0002-9372-5056]{Nour Skaf}
\affiliation{Department of Astronomy \& Astrophysics, University of California, Santa Cruz, CA 95064, USA }

\author[0000-0002-9962-132X]{Ben J. Sutlieff}
\affiliation{Scottish Universities Physics Alliance, Institute for Astronomy, University of Edinburgh, Blackford Hill, Edinburgh EH9 3HJ, UK} \affiliation{Centre for Exoplanet Science, University of Edinburgh, Edinburgh EH9 3HJ, UK}

\author[0009-0001-0275-7811]{Evelyn L. Bruinsma}
\affiliation{Department of Physics \& Astronomy, Johns Hopkins University, 3400 N. Charles Street, Baltimore, MD 21218, USA}

\author[0000-0001-8074-2562]{Sasha Hinkley}
\affiliation{Department of Physics and Astronomy, University of Exeter, Stocker Road, Exeter EX4 4QL, UK}

\author[0000-0002-9803-8255]{Kielan Hoch}
\affiliation{Space Telescope Science Institute (STScI), 3700 San Martin Drive, Baltimore, MD 21218, USA}

\author[0009-0005-6943-6819]{Andrew D. James}
\affiliation{Department of Physics and Astronomy, University of Exeter, Stocker Road, Exeter EX4 4QL, UK}

\author[0009-0009-1074-3696]{Rohan Kane}
\affiliation{Space Telescope Science Institute (STScI), 3700 San Martin Drive, Baltimore, MD 21218, USA}


\author[0000-0002-8895-4735]{Dimitri Mawet}
\affiliation{Cahill Center for Astronomy and Astrophysics, California Institute of Technology, 1200 E. California Boulevard, MC 249-17, Pasadena, CA 91125, USA}
\affiliation{Jet Propulsion Laboratory, California Institute of Technology, 4800 Oak Grove Dr., Pasadena, CA 91109, USA}

\author[0000-0003-1227-3084]{Michael R. Meyer}
\affiliation{Department of Astronomy, University of Michigan, Ann Arbor, MI 48109, USA}

\author[0000-0001-5053-2660]{Skyler Palatnick}
\affiliation{Department of Physics, University of California, Santa Barbara, CA 93106, USA}

\author[0000-0002-3191-8151]{Marshall D. Perrin}
\affiliation{Space Telescope Science Institute (STScI), 3700 San Martin Drive, Baltimore, MD 21218, USA}

\author[0000-0003-2259-3911]{Shrishmoy Ray}
\affiliation{School of Mathematics and Physics, University of Queensland, St Lucia, QLD 4072, Australia}

\author[0000-0003-4203-9715]{Emily Rickman}
\affiliation{European Space Agency (ESA), ESA Office, Space Telescope Science Institute (STScI), 3700 San Martin Drive, Baltimore, MD 21218, USA}

\author[0000-0002-1838-4757]{Aniket Sanghi}
\altaffiliation{NSF Graduate Research Fellow}
\affiliation{Cahill Center for Astronomy and Astrophysics, California Institute of Technology, 1200 E. California Boulevard, MC 249-17, Pasadena, CA 91125, USA}

\author[0009-0008-2252-7969]{Klaus Subbotina Stephenson}
\affiliation{Department of Astronomy \& Astrophysics, University of California, Santa Cruz, CA 95064, USA }



\begin{abstract}

The young M-star TWA 7 hosts a bright and near face-on debris disk, which has been imaged from the optical to the submillimeter. The disk displays multiple complex substructures such as three disk components, a large dust clump, and spiral arms, suggesting the presence of planets to actively sculpt these features. The evidence for planets in this disk was further strengthened with the recent detection of a point-source compatible with a Saturn-mass planet companion using JWST/MIRI at 11 $\mu$m, at the location a planet was predicted to reside based on the disk morphology. In this paper, we present new observations of the TWA 7 system with JWST/NIRCam in the F200W and F444W filters. The disk is detected at both wavelengths and presents many of the same substructures as previously imaged, although we do not robustly detect the southern spiral arm. Furthermore, we detect two faint potential companions in the F444W filter at the 2-3$\sigma$ level. While one of these companions needs further followup to determine its nature, the other one coincides with the location of the planet candidate imaged with MIRI, providing further evidence that this source is a sub-Jupiter mass planet companion rather than a background galaxy. Such discoveries make TWA 7 only the second system, after $\beta$ Pictoris, in which a planet predicted by the debris disk morphology has been detected.

\end{abstract}



\section{Introduction} \label{sec:intro}
Over the past 15 years, continuous improvements to high-contrast imaging (HCI) instruments and techniques have allowed us to directly image more exoplanets with smaller masses and separations from their host star (e.g. 51 Eri b, \citealt{Macintosh15} and AF Lep b, \citealt{DeRosa23,Mesa23,Franson23}). Instruments such as the Gemini Planet Imager (GPI, \citealt{Macintosh14}) on Gemini South and the Spectro-Polarimetric High-contrast Exoplanet REsearch (SPHERE, \citealt{Beuzit19}) on the Very Large Telescope have been pivotal in the direct detection of exoplanets, reaching as far down in mass as $\sim$2-3 M$_{\rm jup}$ \citep{Nielsen19,Vigan21}. This has allowed for the characterization of a new population of young, wide-orbit planets. However, in order to better understand the full demographics of wide-orbit planets ($\gtrsim$10 au), further sensitivity is required to directly image companions of 1 Jupiter mass and below.

In lieu of a HCI instrument capable of imaging sub-Jupiter mass planets, one method of probing these planets has been to indirectly infer them from circumstellar disk substructures such as gaps, warps, and spiral arms. This is difficult to do with protoplanetary disks, as they are gas rich and optically thick.  However, debris disks are optically thin, often gas poor, dusty disks found around main sequence stars, which more clearly bear the imprint of circumstellar disk substructure driven by the presence of planets. These disks proceed the protoplanetary disk phase, and are created and sustained through collisions between planetesimals in the disk \citep{Wyatt08,Matthews14}. In order for planetesimals to collide and produce the dust grains observed, their orbits need to be perturbed, i.e. ``stirred". This requires a stirring mechanism, and while planetesimals may be able to stir themselves \citep{KB08,KB18}, in many cases this requires an unusually massive disk \citep{Krivov21,Pearce22}. Therefore, it is thought that planets are a likely mechanism for stirring these disks and causing collisions \citep{MW09,Meshkat17}.  

Additionally, debris disks harbor complex morphologies, substructures, and asymmetries such as inner clearings, gaps, eccentric disks, warps, spiral arms, radial and brightness asymmetries, etc (e.g. \citealt{Hughes18}), potentially sculpted by unseen planets in these disks. For example, simulations have shown that even a single, 10 Earth mass planet on an eccentric orbit is capable of producing many of the morphologies observed in actual disks \citep{LC16} such as HD 61005 (e.g. ``the Moth", \citealt{Hines07}) and HD 15115 (e.g. ``the Needle", \citealt{Kalas07}). Other examples include the famous $\beta$ Pictoris debris disk, which harbors a warp/secondary disk component, initially predicted to be caused by a planet on an inclined orbit relative to the disk \citep{Mouilett97,Augereau01}. This prediction was confirmed with the detection of the 7-12 Jupiter mass \citep{Brandt21} companion, $\beta$ Pic b, over a decade later \citep{Lagrange10}. More recent work has constrained planet masses and orbits from debris disk properties such as gaps \citep{Marino18}, disk inner edges \citep{Pearce22}, clumps \citep{Skaf23}, and other complex structures such as warps and ``forks" (bifurcation structures due to secondary disk components, \citealt{Crotts24b}), finding that many of these suspected planets are likely sub-Jupiter in mass. These examples demonstrate how debris disk morphologies can be used to infer unseen planetary companions.

Although the direct imaging of sub-Jupiter mass planets has remained difficult due to limitations in HCI, new observatories/instruments are increasing the probability of detecting such planets. The James Webb Space Telescope (JWST, \citealt{Rigby23}), launched in late 2021, is currently the most sensitive observatory for the direct imaging of planet companions in terms of planet mass \citep{Carter23}. Particularly, The NIRCam near-infrared camera yields sensitivity to sub-Jupiter mass planets \citep{Girard22}. Such an instrument is great for planet searches around young, nearby systems, many of which host debris disks. This creates an opportunity to study the population and architectures of small, wide-orbit planets, in addition to probing planet-disk interactions and exoplanetary system evolution.

In this paper, we present NIRCam observations of the debris disk system, TWA 7 (CE Ant), in the F200W and F444W filters. TWA 7 is a young ($\sim$7-13 Myr, \citealt{Bell15,Binks20,Luhman23}) M-dwarf in the TW Hydra association (TWA), at a distance of 34 pc \citep{Gaia20}. What makes this system so interesting is its near face-on and complex debris disk. The first published detections of the disk include photometric measurements of the infrared excess present in the spectral energy distribution (SED) with Spitzer \citep{Low05} and the SCUBA instrument on the James Clerk Maxwell Telescope \citep{Matthews07}. Additionally, the disk has been spatially resolved over the last three decades with instruments such as NICMOS and STIS on the Hubble Space Telescope  (HST; \citealt{Choquet16, Ren21}), SPHERE on the Very Large Telescope \citep{Ren21}, and the GPI on Gemini South \citep{Esposito20}. The disk has been shown to host several different substructures, including three distinct disk components, spiral arms, and a southern dust clump \citep{Olofsson18,Ren21}. The three disk components consist of a bright, relatively broad inner ring, a very narrow ring outside the inner ring, and a very broad, faint outer disk. The narrow, secondary ring has been hypothesized to be a possible resonant structure caused by an unseen planet companion \citep{Ren21}. If true, this planet was predicted by \citet{Ren21} to reside in an under-dense region of the ring in the northwest at 1.5$''$ from the star. Recently, JWST/MIRI 11$\mu$m observations revealed a bright candidate companion at the location of the predicted planet \citep{Lagrange25}. If planetary in nature, the companion would have the mass of Saturn. This would make the planet companion the lowest mass planet directly imaged to date, as well as only the second system, after $\beta$ Pictoris, where the discovered planet was originally predicted based on its debris disk morphology. Given the complex debris disk structures and a planet candidate directly interacting with the disk, this makes TWA 7 a truly unique and important system to study.

In Section \ref{sec:obs}, we discuss our NIRCam observations and data reduction methods and techniques. We then describe in Section \ref{sec:mods} our forward modeling procedure, which allows us to effectively subtract a disk model from the data and increase contrast at those separations. In Section \ref{sec:disk} we discuss the disk morphology as seen by NIRCam and compare these results to previous observations of the disk. Finally, in Sections \ref{sec:companions} and \ref{sec:planets}, we discuss our search for planetary companions in our observations and provide results from a companion extraction procedure. This includes a 2-3$\sigma$ detection at the same location as the companion observed with MIRI (further discussed in Section \ref{sec:c6}) and two other possible companions found both outside and embedded within the disk.

\begin{table*}
	\centering
	\caption{\label{tab:data_sum}Summary of the data used in this paper. Here, $N_{ints}$ is the total number of integrations, $N_{groups}$ is the number of groups per integration, and $N_{frames}$ is the number of frames per group. Additionally, t$_{int}$ is the effective integration time in seconds and t$_{exp}$ is the total effective exposure time in seconds.}
	\begin{tabular*}{\textwidth}{c @{\extracolsep{\fill}} cccccc}
	    \hline
	    \hline
		Name & Filter & Date & Readout Pattern & $N_{ints}$/$N_{groups}$/$N_{frames}$ & t$_{int}$ (s) & t$_{exp}$ (s) \\
		\hline
		  \multirow{ 2}{*}{TWA 7} & F200W & 2024 Jun 15 & DEEP8 & 5/18/8 & 372 & 1860  \\
		& F444W & 2024 Jun 15 & DEEP8 & 5/18/8 & 372 & 1860  \\
		\hline
		\hline
	\end{tabular*}
\end{table*}

\begin{figure*}
    \centering
    \includegraphics[width=\textwidth]{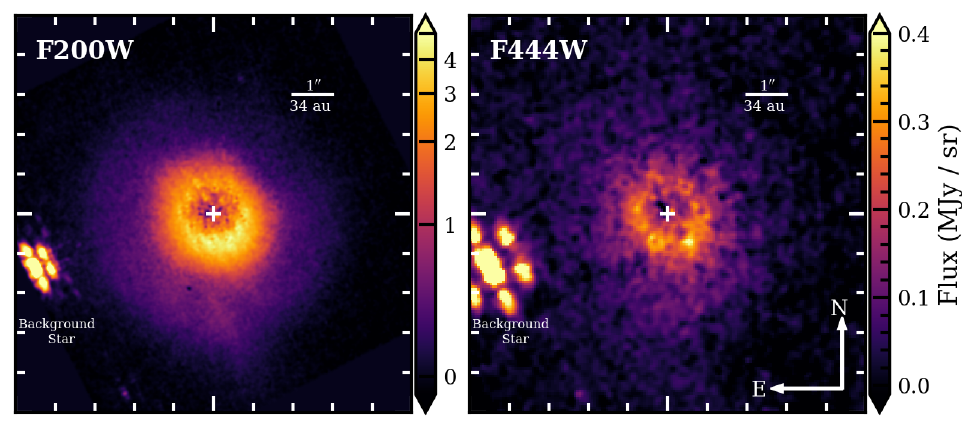}
    \caption{\label{fig:reduc} Final MCRDI data reductions for TWA 7 in the F200W (\textit{left}) and F444W (\textit{right}) filters. Both images are rotated with North up, and the $+$ represents the star location. Each tick represents 1$''$ ($\sim$34 au), with a total FOV of 10$''$ by 10$''$. Additionally, the surface brightness for the F200W data is scaled linearly for $<$1 MJy/sr and is in log scale for $>$1 MJy/sr to better show the complex disk morphology.} 
\end{figure*}

\section{Observations \& data reductions} \label{sec:obs}

\subsection{Observations}
TWA 7 was observed on June 15 2024 UT with the JWST/NIRCam Coronagraphy mode \citep{Girard22} as part of the survey ``Uncharted Worlds: Towards a Legacy of Direct Imaging of Sub-Jupiter Mass Exoplanets" (GO 4050, PI: A. Carter). The goal of this survey is to target a sample of young, nearby systems in the TWA association in search of new sub-Jupiter mass companions. This included 23 science targets and 7 dedicated references. TWA 7 was imaged simultaneously (dual-channel mode) in the F200W ($\lambda_{\text{pivot}} = 1.99 \ \mu$m, $\Delta \lambda = 0.46 \ \mu$m) and the F444W ($\lambda_{\text{pivot}} = 4.40 \ \mu$m, $\Delta \lambda = 1.02 \ \mu$m) filters and using the intermediate round occulter MASK335R with an inner working angle of $0\farcs63$. These two filters were chosen specifically as planets should be significantly brighter at 4 $\mu$m compared to 2 $\mu$m (e.g. \citealt{Skemer14}), therefore, the color measured between the F200W and F444W filters can help discriminate between a potential planet candidate and a background object. The F200W observation was taken with a field of view (FOV) of $10\arcsec \times 10\arcsec$ and a pixel scale of 31 mas per pixel, while the F444W observation was taken with a FOV of $20\arcsec \times 20\arcsec$ and a pixel scale of $\sim$63 mas per pixel. Table \ref{tab:data_sum} displays the summary of the observations for both filters.

\subsection{Data reduction}
All data were reduced using the codes \texttt{spaceKLIP}\footnote{\url{https://github.com/spacetelescope/spaceKLIP}} \citep{Kammerer22,Carter23} and \texttt{Winnie}\footnote{\url{https://github.com/kdlawson/Winnie}} \citep{Lawson22}. The unprocessed science and reference images are first read into the \texttt{spaceKLIP} reduction pipeline, where they undergo three stages of reduction. In Stage 1, the pipeline first applies detector-level corrections for saturation, superbias, non-linearity, and dark current, followed by ramp-fitting of the raw, uncalibrated data. We note that during this stage, we turn off the 1/f noise correction as removing this noise can lead to artifacts in NIRCam images specifically with disks. The stage 1 corrected data are then further processed in stage 2 of the pipeline, where additional correction and calibration are performed, including background subtraction and flat-fielding. Finally, the corrected data are prepared for post-processing in Stage 3. In this stage, the data are median-subtracted, bad-pixel repaired, improved in PSF centering and alignment.

The next step is subtracting the stellar PSF from the observations through Reference star Differential Imaging (RDI). To maximize the quality of the PSF subtraction, a large reference PSF library was put together by compiling all observations from the GO 4050 survey. A large number of reference stars provides useful diversity in the PSFs, resulting in a better PSF subtraction (e.g. \citealt{Choquet16,Xuan18,Xie22,Aniket22,Aniket24}). At the same time, all stars in the survey have similar spectral types (most are M-stars) and have minimal wavefront drift between them, as most were taken within a few days time span, making them an ideal set of reference stars. 

We further identify reference targets that should not be included in the PSF library. All stars in the survey were themselves PSF-subtracted first using the PSF library to identify targets with disks, with sources hiding behind the PSF or with nearby overly bright sources. These observations were excluded from the library to avoid contamination of our science target. Two additional observations were excluded from the library due to bad integrations (possible tilt events during observation). Several of the remaining observations had faint sources in the field, such as background stars or galaxies. To prevent these from contaminating the subtraction, small circular regions centered at these sources were defined in each observation (with radii corresponding to the size of each specific source), and the pixels in all of these circles were substituted with the stellar flux-weighted median value of that pixel across all observations in the library. From a total of 91 dual-filter observations in the GO 4050 program spanning 30 stars (of which the 7 dedicated references were observed with a 9-POINT-CIRCLE dither pattern and a few of which had repeated observations), 81 observations for the F200W filter and 83 observations for the F444W filter were chosen to constitute the PSF library.

With the PSF library, the corrected and cleaned TWA 7 data are ready for post-processing. Instead of using \texttt{spaceKLIP} which uses Karhunen-Lo\'eve image processing (KLIP, \citealt{Soummer12}) for performing RDI, we instead run the reduced datasets through the code \texttt{Winnie}. This code is chosen as \texttt{Winnie} is a \texttt{spaceKLIP} compatible code designed specifically for JWST coronagraphic disk observations. Once the data are read into a database, using the reference images, \texttt{Winnie} performs RDI on the science images to produce a single PSF-subtracted image of the target. Model-Constrained RDI (MCRDI; \citealt{Lawson22}) is then performed to optimize our PSF-subtracted images and minimize the loss of disk signal. This process involves forward-modeling and using best-fit synthetic disk models to fine-tune RDI constraints and create a higher signal-to-noise (SNR) reduction. Further details on the forward-modeling process will be discussed in Section \ref{sec:mods}. Finally, MCRDI images are smoothed using a Gaussian filter with $\sigma$ = 1 pixel, to further increase the SNR. Our final images of TWA 7 in both filters can be seen in Figure \ref{fig:reduc}. We note that there is an overestimation in the background subtraction for the F200W data, causing the background level to be slightly negative. We account for this overestimation during the forward-modeling procedure as explained in Section \ref{sec:procedure}.

\section{Disk forward modeling} \label{sec:mods}

\begin{figure*}
    \centering
    \includegraphics[width=\textwidth]{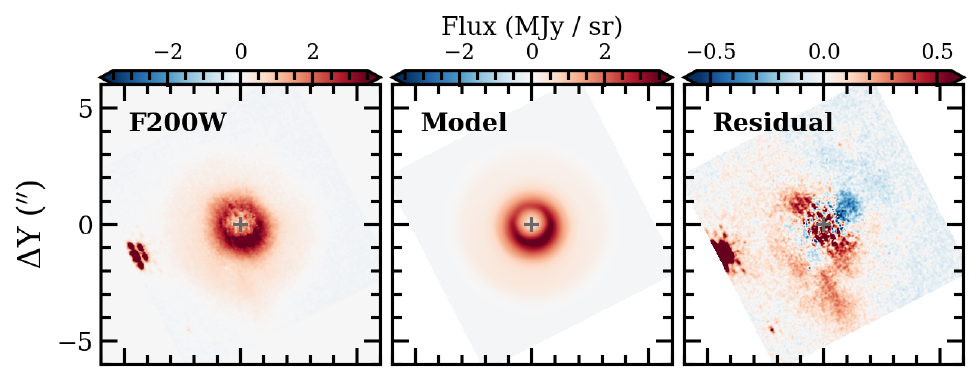}
    \includegraphics[width=\textwidth]{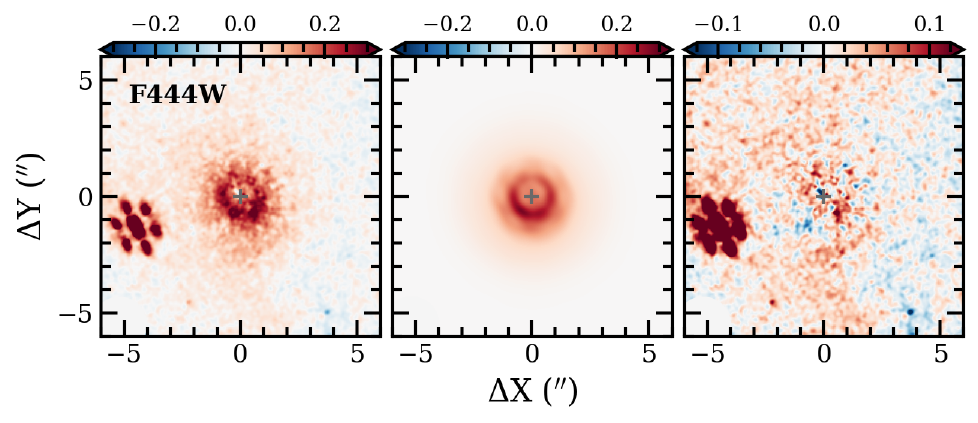}
    \caption{\label{fig:mod2micron} Forward modeling results for TWA 7 in the F200W filter (\textit{top}) and the F444W filter (\textit{bottom}). Left is our NIRCam observations, middle is the best fitting forward model, and right is the resulting residual map. All images are rotated with North up, and the $+$ represents the star location.} 
\end{figure*}

\subsection{Procedure} \label{sec:procedure}
Once all the data are cleaned and ready for RDI, \texttt{Winnie} is then used to perform forward modeling. The goal of this process is to generate MCRDI images, as well as to find a best-fit disk model to subtract from the data and increase contrast at those separations for planet searches and to highlight any disk substructures present. For our model, we use a GRaTer model, i.e. a simple ring-like disk morphology based on \citet{Augereau99} which assumes a 2-parameter Henyey-Greenstein (HG; \citealt{HG41}) scattering phase function (SPF). The models are defined using several different parameters: The fiducial disk radius ($r_{0}$), the ratio between the scale height at $r_{0}$ and $r_{0}$, i.e. $h(r_{0})/r_{0}$ ($h_{0}$), inclination ($\theta_{inc}$), position angle ($\theta_{PA}$), the radial density power law exponent interior and exterior to $r_{0}$ ($\alpha_{in}$ and $\alpha_{out}$, respectively), the vertical density exponent ($\gamma$), two HG asymmetry parameters ($g_{1}$ and $g_{2}$, respectively), and the weight for the SPF with the asymmetry parameter $g_{1}$ ($w_{1}$). Once the model is created, it is forward modeled and convolved with the PSF to match the observations. 

To fit our models to the data, we simply use a minimization routine via the code \texttt{lmfit.minimize} \citep{lmfit25} to find the best-fitting parameters. We choose the Powell minimization method in order to reduce computational time while still being able to find models that are a good fit to the data. We first start the minimization procedure using initial values for each parameter. These initial values were chosen based on modeling results from \citet{Ren21}, who also use a GRaTer model to fit each disk component observed in the SPHERE data. Similar to \cite{Ren21}, we model all three disk components in order to achieve the best overall fit. To reduce parameter space and computational time, we allow some parameters to vary while some remain fixed. This includes parameters such as inclination, which is already well-constrained from previous modeling. In addition to the disk structure and SPF parameters, we also include a relative flux parameter, $F$. The $F$ parameter is fixed at 1 for the brightest inner ring component, while varying for the outer two disk components in order to fit the relative flux of these two disk components compared to the inner ring component. Finally, to address an overestimation in the background subtraction for the F200W data, we include a background level parameter. 

During the minimization routine, roughly 10,000 models are produced within a range of priors for each parameter, where an optimized model is then returned. The models are fit specifically to the F200W data, which has a much stronger detection of the debris disk. The best-fit values for each parameter are then used to create a forward model for the F444W observation. In order to save computational time, we do not calculate uncertainties for each parameter, as the main goal is not to further constrain these disk parameters, given that the disk has been extensively modeled in previous works (e.g. \citealt{Olofsson18,Esposito20,Ren21,Crotts24}) and because of the fact that we are using a simple model to fit a complex disk.  Rather, as mentioned previously, our goal is to fit the disk flux in order to subtract the model from our data to improve contrasts for planet searches and to analyze any underlying disk substructure. 

\subsection{Results}
Results for the forward modeling can be found in Table \ref{tab:mod_res}, where each row shows the minimized value found for each parameter. We include the results for each disk component, with Ring 1 being the inner ring, Ring 2 being the narrow secondary ring, and Ring 3 being the faint, broad outer disk component. Figure \ref{fig:mod2micron} shows our MCRDI reductions for both filters in the left panels (F200W, top, and F444W, bottom), while the best-fitting model is shown in the middle panels, and finally the residuals after subtracting the model from the data are shown in the right panels. Although our best-fit model does a decent job of matching the disk flux in both filters, significant residuals remain for the F200W filter due to the complex substructures present in the disk. These substructures will be discussed further in the following section.

\begin{table}
	\centering
	\caption{\label{tab:mod_res}List of best fitting model parameters for each disk using a minimization routine. Parameters which were fixed during the minimization process are labeled.}
	\begin{tabular}{cccc}
	    \hline
	    \hline
		Parameter & Ring 1 & Ring 2 & Ring 3 \\
		\hline
		  $R_{0}$ (au) & 27.11 & 51.4 (fixed) & 104.90 \\
		$\theta_{inc}$ ($^{\circ}$) & 13 (fixed) & ... & ...  \\
            $\theta_{PA}$ ($^{\circ}$) & 88.0 & 94.0 & 77.42  \\
            $h_{0}$ & 0.05 (fixed) & ... & ... \\
            $\alpha_{in}$ & 4.45 & 12 (fixed) & 1.73 \\
            $\alpha_{out}$ & -4.21 & -15 (fixed) & -15 \\
            $\gamma$ & 2.98 & 1.44 & 1.75 \\
            $g_{1}$ & 0.78 & 0.99 & 0.94 \\
            $g_{2}$ & 0.82 & 0.90 & 0.88 \\
            $wg_{1}$ & 0.66 & 0.52 & 0.83 \\
            $F$ & 1 (fixed) & 0.79 & 0.41 \\ 
		\hline
            Background Level \\ (MJy/sr) & -0.06 & & \\
            \hline
	\end{tabular}
\end{table}

\section{Disk Analysis} \label{sec:disk}

\subsection{Observed substructures with NIRCam}
As mentioned previously, the TWA 7 disk is known to harbor several different substructures. This includes three distinct disk components, spiral arm structures, and a southern dust clump \citep{Choquet16,Olofsson18,Ren21}. All three disk components are seen in the majority of observations of the disk taken in scattered light, including NICMOS \citep{Choquet16}, SPHERE, and STIS \citep{Olofsson18,Ren21}  (the GPI FOV is too small to see the outer two rings). The southern dust clump is also detected with NICMOS, SPHERE, STIS, and possibly ALMA in the sub-millimeter which presents extended emission in the same location \citep{Bayo19}, although the disk is only marginally detected. The spiral arm is detected with SPHERE and STIS in 2017 and 2019, while only marginally detected with GPI in 2018 \citep{Esposito20,Crotts24}.

\begin{figure}
    \centering
    \includegraphics[width=0.472\textwidth]{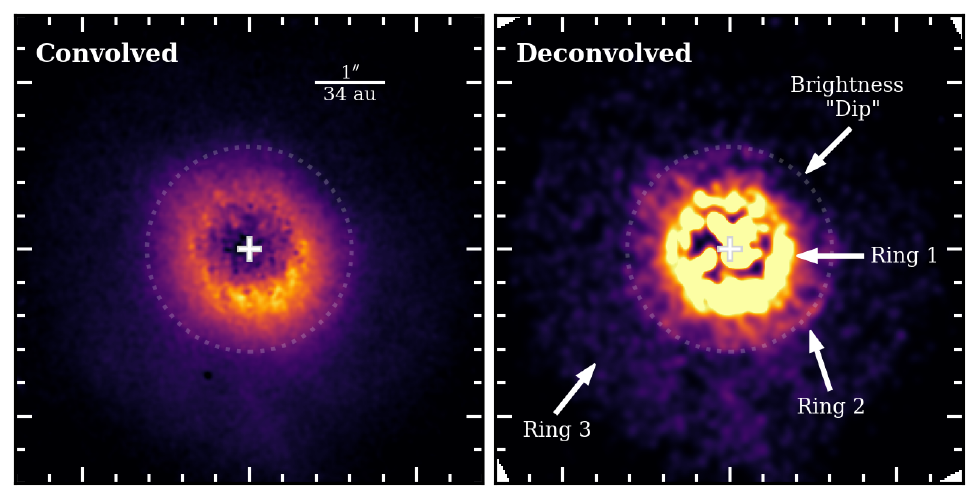}
    \includegraphics[width=0.472\textwidth]{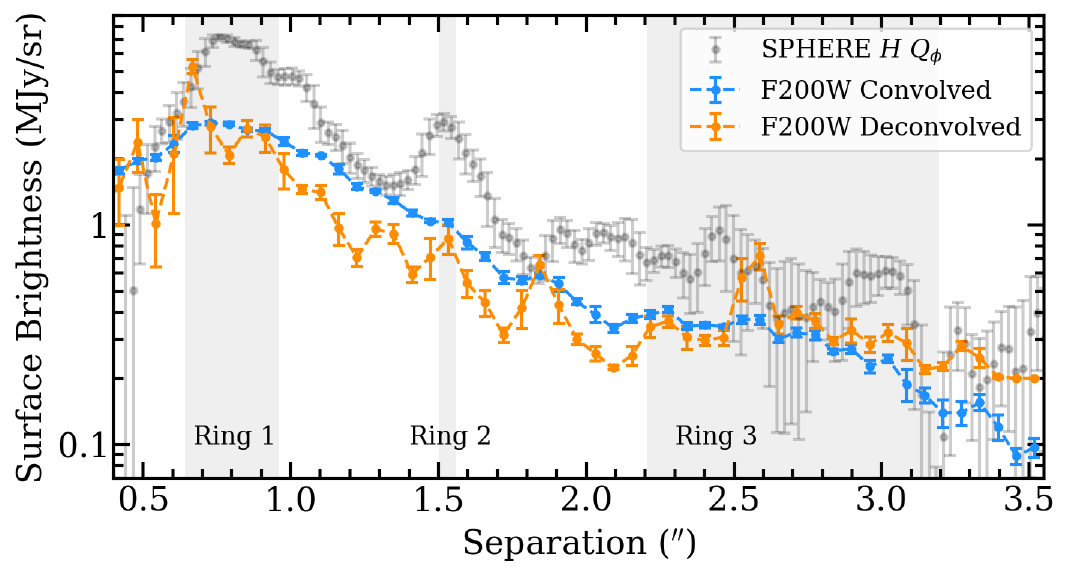}
    \caption{\label{fig:deconv_twa7} \textbf{Top Left:} MCRDI reduction of TWA 7 in the F200W filter. \textbf{Top Right:} The same reduction as in the left panel, but deconvolved. Arrows and labels point to the three ring components that make up the disk, as well as the dip in surface brightness observed in the northwest region of Ring 2. The dashed circle represents the location of the second ring component. \textbf{Bottom:} Radial surface brightness profile of the TWA 7 disk between our convolved (light blue) and deconvolved (orange) F200W images compared to the 2017 SPHERE data (black). The three ring locations are highlighted.} 
\end{figure}

For our NIRCam observations, we find the disk to be present in both the F200W and F444W filters, with the disk being significantly brighter in the F200W filter. As mentioned in Section \ref{sec:mods}, we fit all three disk components during our forward modeling procedure as all three are present in our data. This being said, the second ring component, Ring 2, is much harder to distinguish. In Figure \ref{fig:deconv_twa7}, we present a deconvolved version of our reductions in the F200W filter, created with \texttt{Winnie}, which allows Ring 2 to become more apparent. To better visualize the disk's radial structure, we also present the disk surface brightness as a function of radial separation from the star. The surface brightness is measured by placing $\sim$0.1$''$ by 0.1$''$ square apertures along a radial slice of the disk (specifically the south side where the disk is brightest). The average surface brightness is measured for each aperture, while the 1$\sigma$ uncertainties are measured by taking the standard deviation of each aperture. We compare the disk's radial surface brightness between the F200W images before and after deconvolution, as well as the SPHERE $H$-band polarized intensity (see bottom panel of Figure \ref{fig:deconv_twa7}). Ring 2 is clearly visible in the SPHERE radial surface brightness profile, while the convolved F200W profile is significantly smoother, with no clear sign of Ring 2. While the discrepancy in the radial profile may be due to differences in photon scattering between polarized and total intensity, another contributing factor may be that the PSF wings from NIRCam are much broader than those of SPHERE. This would explain why deconvolving the disk makes Ring 2 more visible, although it is still much fainter and not as well defined as in polarized intensity. Regardless, we can still confirm that Ring 2 is at a consistent distance compared to previous observations of $\sim$1.53$''$ or 52.2 au. For the F444W filter, only Ring 1 is significantly detected, while Ring 2 and 3 are not significantly detected.

\begin{figure}
    \centering 
    \includegraphics[width=0.47\textwidth]{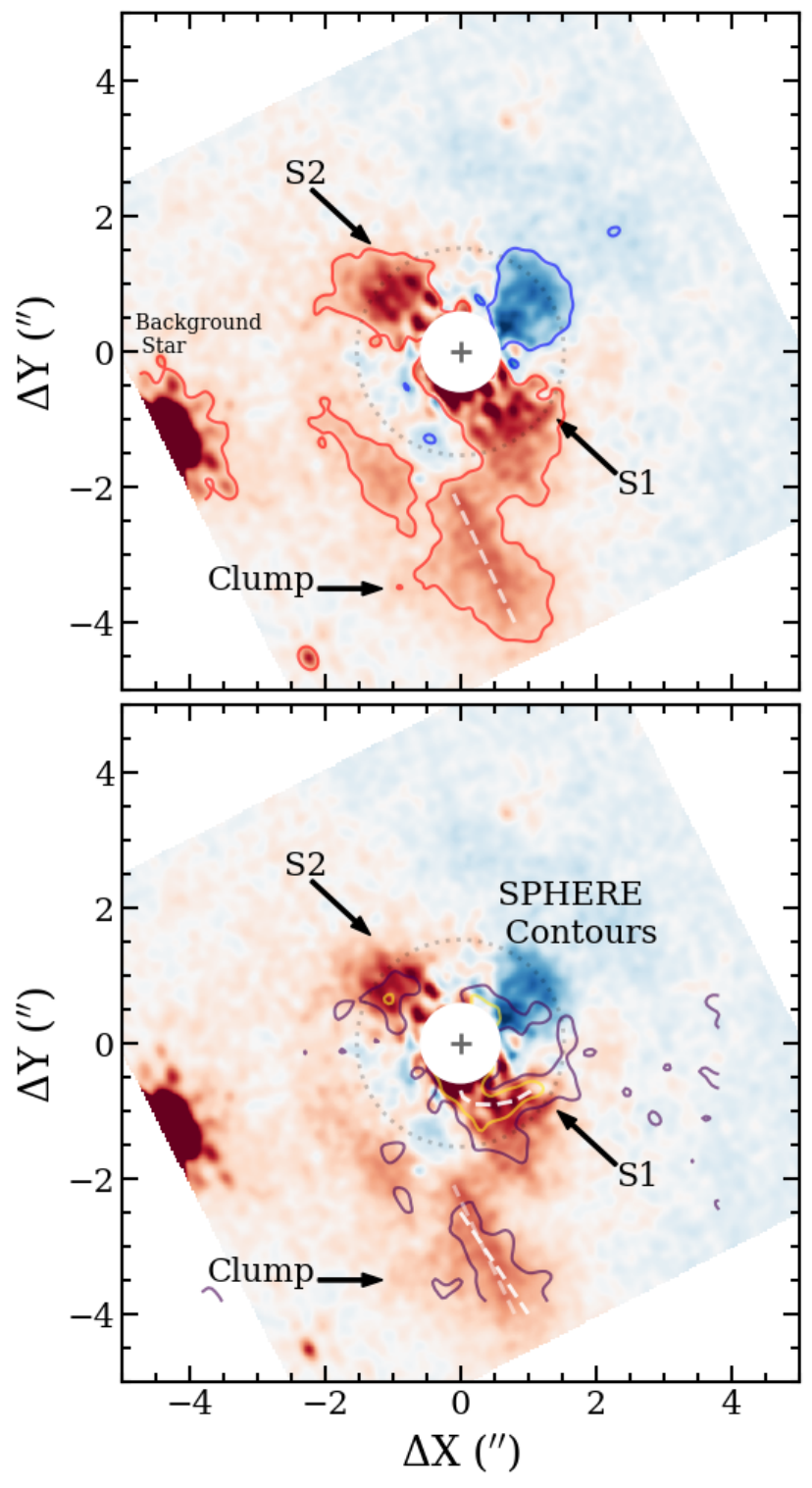}
    \caption{\label{fig:twa7} Residual maps of TWA 7 in the F200W scaled between -0.5 and 0.5 MJy/sr. \textbf{Top:} Red contours represent a surface brightness of 0.12 MJy/sr and blue contours represent a surface brightness of -0.12 MJy/sr in the F200W to highlight the positive and negative residuals that remain after the disk model subtraction. The lower left dashed line represents the location of the dust clump as observed by NIRCam. \textbf{Bottom:} Contours represent a surface brightness of 7 and 20 $\mu$Jy arcsec$^{-2}$ (0.30 and 0.85 MJy/sr) in the model-subtracted SPHERE data to highlight the location of the spiral arms and dust clump \citep{Ren21}. The lower left dashed lines represent the location of the dust clump as observed by NIRCam (transparent) and SPHERE (opaque), while the curved dashed line just south of the star represents the location of the spiral arm as observed with SPHERE. We label the residuals similar to \citet{Ren21}, where S1 and S2 are the south and north spiral arms observed with SPHERE and Clump refers to the southeast dust clump.}
\end{figure}

After subtracting our best-fitting model from the F200W data, we find multiple significant residuals which are mostly consistent with substructures seen in other observations. This includes significant residuals just north and south of the star, as well as southeast of the star. These substructures are highlighted by the contours in the top panel of Figure \ref{fig:twa7}. Given the greater sensitivity of our observations, further substructures can be seen in the F200W residuals. For example, previous observations with HST and SPHERE show the southern clump to be isolated from the inner disk. However, our F200W observations show that the southern clump appears to be connected to the inner disk component between the clump and the southern spiral arm. Further significant residuals are seen above the southern clump to the east, suggesting either a second disk clump, or a further extension of the southern disk clump. Comparing the position of the clump to SPHERE observations, represented by the contours in the bottom panel of Figure \ref{fig:twa7}, shows that the clump has not moved significantly since 2017, although it may have rotated slightly as indicated by the dashed lines. While we would expect the clump to have rotated 2$^{\circ}$-4$^{\circ}$ over the course of seven years depending on the separation from the star (assuming a Keplarian rotation and host mass of 0.46 M$_{\odot}$), the large width of the clump, as well as the lower sensitivity of SPHERE, make it difficult to measure such a small rotation and therefore make any strong conclusions about whether or not the clump has indeed rotated. 

At the location of S1 and S2, we do not find clear evidence for spiral arms, but rather there are two groups of clumpy residuals north and south of the star. Comparing S1 and S2 between SPHERE and NIRCam, S2 appears relatively the same with NIRCam, although slightly more extended to the north. In contrary, while S1 appears as a distinct spiral arm in the SPHERE data, we again only see clumpy residual substructure at the same location with NIRCam. There is a residual just south of the SPHERE S1 location that almost resembles a spiral arm, however, given the separation from the star, this turns out to only be a residual of Ring 2. Further SPHERE observations therefore may be necessary to confirm that S1 is still present, and if so, whether or not it has significantly changed.

\subsection{Azimuthal surface brightness of Ring 2} \label{sec:ring2}
In addition to comparing the spiral arm and southern clump structures, we also take the opportunity to compare the azimuthal surface brightness profile of the narrow Ring 2 between our F200W data and previous observations. This is of interest, as Ring 2 has been shown to harbor an under-density or ``dip" in surface brightness in the northwest region of the ring (see right panel of Figure \ref{fig:deconv_twa7}), which is where a planet was predicted to reside \citep{Ren21} and is also the location where the planet candidate has been detected in MIRI observations \citep{Lagrange25}. Comparing the location of this surface brightness dip between the NICMOS, SPHERE, and STIS observations, \citet{Ren21} found that this region of the ring appeared to shift $\sim 19\fdg8 \pm 1\fdg0$ per year between 1998 and 2019. The explanation for such a large rotation is that the dip may be a shadowing effect from a source orbiting closer to the star at $\sim$5 au. 

\begin{figure}
    \centering
    \includegraphics[width=0.472\textwidth]{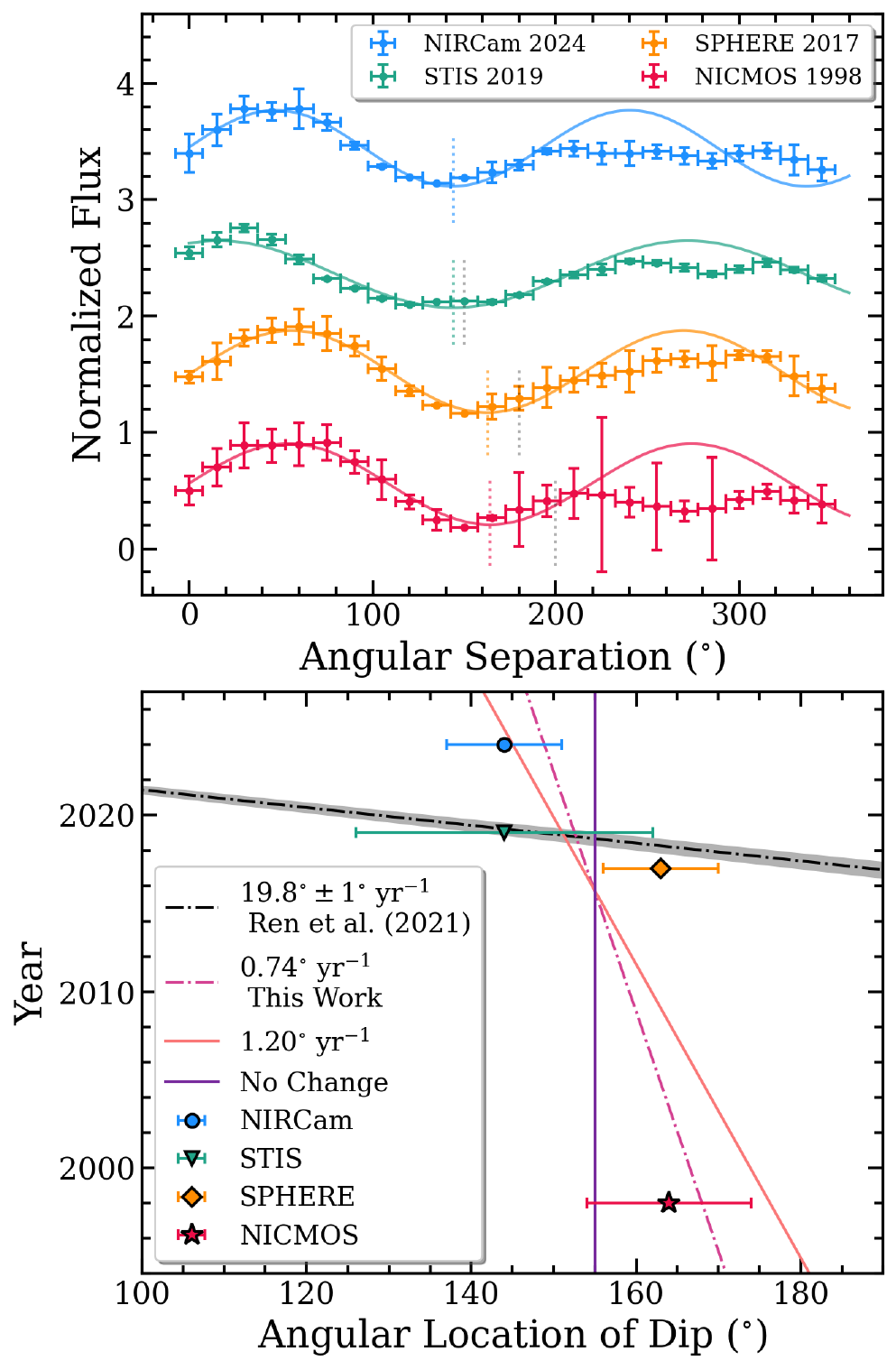}
    \caption{\label{fig:sb_ring2} \textbf{Top:} Azimuthal surface brightness of Ring 2 as a function of angle in degrees normalized between 0 and 1, although we vertically separate each profile for easier viewing. The surface brightness is measured for our F200W data, as well as the 1998 NICMOS, 2017 SPHERE and 2019 STIS data from \citet{Choquet16} and \citep{Ren21}. Additionally we include the best fit cosine curves and colored dashed lines representing the measured dip location. The grey dashed lines represent the dip location measured in \citet{Ren21} for reference. \textbf{Bottom:} Year each dataset were taken, versus the angular separation of the dip in degrees. The black dashed line and grey shaded region represents the fit with uncertainties to the data measured in \cite{Ren21}. The three colored lines represent the following scenarios; 1) no change in the dip location (solid purple), 2) a rotation of $0\fdg74$ yr$^{-1}$ based on the best fit linear model to all four data point (dashed pink), and 3) a rotation of $1\fdg20$ yr$^{-1}$ based on the expected rotation of Saturn mass planet at $\sim1\farcs5$ (solid pink-gold). All errorbars represent 1$\sigma$ uncertainties.} 
\end{figure}

Using the same publicly available NICMOS, SPHERE and STIS observations from \citet{Choquet16} and \citet{Ren21}, we conduct the same analysis and additionally include our F200W NIRCam observations to see if the location of the under-dense region of Ring 2 fits the same trend. We start by placing 24, $\sim 0\farcs15$ by $0\farcs15$ sized square apertures (taking into account the different pixel scales between instruments) along Ring 2 between 0 and 360 degrees, therefore placing each aperture $15^{\circ}$ apart. We define an angle of $0^{\circ}$ as the center of the southern side of Ring 2 and measure counterclockwise around the ring. For each aperture, we measure the average disk flux and the standard deviation. Finally, the average flux is normalized (between 0 and 1) and plotted against the angular separation, as can be seen in the top panel of Figure \ref{fig:sb_ring2}. To determine the angular location of the Ring 2 surface brightness dip, we fit a cosine function to the measured azimuthal surface brightness similar to \citet{Ren21}. Since the backside of the disk is fainter than the front side due to scattering effects, we find that fitting the cosine function only out to 200$^{\circ}$ achieves a better fit at the dip location. This results in a dip location of $156^{\circ} \pm 7^{\circ}$, $156^{\circ} \pm 3^{\circ}$, $139^{\circ} \pm 15^{\circ}$, and $144^{\circ} \pm 7^{\circ}$, for NICMOS, SPHERE, STIS and NIRCam, respectively. The 1$\sigma$ uncertainties are derived based on the fit of the cosine model to the data.

To measure the rotation rate of the dip in Ring 2, we plot the angular location of the dip for all three observations versus the year the data were taken (1998, 2017, 2019, and 2024) shown in the bottom panel of Figure \ref{fig:sb_ring2}. We use \texttt{scipy odr} to estimate linear fits to our four data points and their uncertainties, yielding a clockwise rotation of the dip of $0\fdg74^{+1.77}_{-0.30}$ per year. We note that there are several factors that may affect our result. This includes the fact that the NICMOS data is fairly low SNR and all four observations vary in wavelength. Additionally, the SPHERE data is in polarized intensity while the other three observations are in total intensity. In either case, this analysis shows that the change in the angular location of the dip or underdense region of Ring 2 is significantly less than the $19\fdg8$ yr$^{-1}$ rotation found in \citet{Ren21}, and therefore is unlikely to be a shadowing effect from a source orbiting closer to the star as suggested by \citet{Ren21}. Even if we assume that the dip location has rotated 360 degrees between 1998 and 2017 (as assumed by \citealt{Ren21} leading to the measured $19\fdg8$ rotation), the dip location observed with NIRCam in 2024 still rules this scenario out, as we would then expect the dip to be on the opposite side of the disk from what is observed. Instead, our results are more consistent with a planetary mass object at that location. Assuming a stellar mass of 0.46 M$_{\odot}$ for the host star \citep{Stassun18} and a semi-major axis of 49.5 au ($1\farcs49$ based on the distance measured from \citealt{Lagrange25}), a Saturn-mass companion would have an orbital motion of $\sim1\fdg2$ per year. While this is almost twice as high as our result of $0\fdg74$ per year, it is still consistent within 1$\sigma$ uncertainties. For reference, we plot a line for $1\fdg2$ yr$^{-1}$ in the bottom subplot of \ref{fig:sb_ring2}, as well as a vertical line which indicates no change in the dip location over time (a scenario consistent within 2.5$\sigma$ uncertainties). This finding gives further evidence for a planetary companion at the location of Ring 2. For a more robust analysis, multiple observations over time using the same instrument and observing mode will be useful to constrain the rotation of the underdense region in Ring 2 even further.

\section{Companion candidates} \label{sec:companions}

\begin{figure*}
    \centering 
    \includegraphics[width=\textwidth]{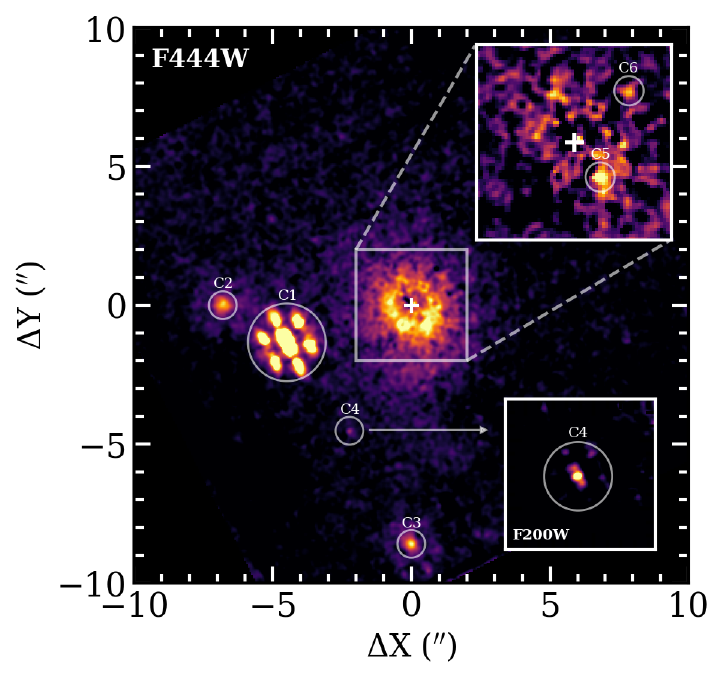}
    \caption{\label{fig:candidates} Full FOV of the TWA 7 observation in the F444W filter. Each candidate is circled and labeled with C1-6. The lower right plot shows a zoom in on C4 in the F200W filter, while the upper right plot shows a zoom in on the inner region of the F444W observation after subtraction of the disk model to better show candidates C5 and C6.}
\end{figure*}

\begin{table*}
	\centering
	\caption{\label{tab:candidate_sum}Summary of the candidates identified in the F444W filter. ($\delta$x,$\delta$y) positions are only rough estimates made by eye. $^{*}$Sources are outside the FOV of the F200W filter}
	\begin{tabular*}{\textwidth}{c @{\extracolsep{\fill}} ccc}
	    \hline
	    \hline
		Name & Estimated Position ($\delta$x,$\delta$y) & Seen in F200W? & Confirmed Background source? \\
		\hline
		  C1 & (-4.49$^{''}$ , -1.34$^{''}$) & Yes & Yes \\
            C2 & (-6.78$^{''}$ , 0.05$^{''}$)  & No$^{*}$ & Yes \\
            C3 & (0.00$^{''}$ , -8.60$^{''}$)  & No$^{*}$ & Yes \\
            C4 & (-2.23$^{''}$ , -4.53$^{''}$) & Yes & No \\
            C5 & (0.53$^{''}$ , -0.70$^{''}$) & No & No \\
            C6 & (1.10$^{''}$ , 1.01$^{''}$)  & No & No \\
		\hline
		\hline
	\end{tabular*}
\end{table*}

These observations have given us the opportunity to further analyze the disk structure and compare to previous observations. However, the main goal of this work is to search for sub-Jupiter mass planet companions in the TWA 7 system. With this in mind, we take a closer look at both the F200W and F444W observations to identify both known and unknown companions present in our images. Although the F444W filter is expected to reach a deeper planet mass sensitivity compared to the F200W filter, as a planet should be much brighter at these longer wavelengths, observations with the F200W will be important for discriminating between planet companions and background objects.

We identify by manual inspection six point-like and extended sources in our images between the F200W and F444W filters. These sources are circled in Figure \ref{fig:candidates} and are consecutively labeled from brightest (C1) to dimmest (C6). Their estimated ($\delta$x,$\delta$y) positions (+$\delta$x is right of the star and +$\delta$y is north of the star), are listed in Table \ref{tab:candidate_sum}. All six sources are observed in the F444W filter, while only two are seen in the F200W filter (C1 and C4), although C2 and C3 are outside the F200W FOV. Two out of the six sources have been previously observed; this includes C1 and C2. C1 is a known background star that has been seen in observations with NICMOS, STIS, and MIRI. Its position remains stationary with the proper motion of TWA 7. C2 is most likely a background galaxy, given that it is spatially extended and is also found to be consistent with the proper motion between our NIRCam observations and observations of the galaxy seen with ALMA in 2016 and MIRI. Similar to C2, C3 is another extended background galaxy that has also been observed with MIRI.

The next two sources, C4 and C5, are not present in any previous observations. C4 is detected in both the F200W and F444W and is located $\sim$5$''$ from the star. C5 is only detected in F444W and is much closer to the star ($\sim$0.9$''$ or 30.7 au) between Ring 1 and Ring 2. While C5 initially appears to be a part of the disk, it cannot be explained by the disk model. Finally, C6 is the faintest (detected at the 2$\sigma$-3$\sigma$ level) but most compelling source, located $\sim$1.5$''$ northwest of the star. Similar to C5, C6 is observed in F444W, but does not appear to have any counterpart in F200W. Although this source is too faint to draw any strong conclusions about its nature via the F444W observations alone, a much brighter source is detected with MIRI at the exact same location \citep{Lagrange25}, which was found to be consistent with a Saturn-mass planet. Assuming C6 is the same source as the MIRI candidate, our NIRCam observations will provide additional information to further characterize this potential planet companion. 

\subsection{Jack-knife test}
Since C4-C6 have not yet been identified as background objects, we focus on these sources as potential planet candidates. However, we first want to confirm that the three candidates are not artifacts introduced by one of the reference images, which can commonly occur for faint objects. For this we use the jack-knife tool in \texttt{Winnie} which repeats PSF-subtraction $N_{ref}$ times, leaving out one reference each time. We can then inspect each PSF subtraction to ensure that the candidates are consistently present in the data. If one of the candidates is introduced into the image due to one of the reference images, this typically results in the candidate being present in only some of the frames, but not all. Performing the jack-knife test on both our F200W and F444W data, we find that all three planet candidates are present in every frame at the same SNR, confirming that they are real sources in the TWA 7 images and not artifacts introduced during the PSF subtraction process.

In the following sections we model each of the three candidates to extract their position and flux. These flux measurements are then used to calculate the color and potential planet mass via planet evolutionary models. 

\subsection{Contrast curves \& companion modeling} 

\subsubsection{Contrast curves} \label{sec:contrast}
To estimate the sensitivity of our observations, we again utilize both the \texttt{spaceKLIP} and \texttt{Winnie} codes to calculate 5$\sigma$ contrast curves. On its own, \texttt{spaceKLIP} calculates contrast curves using KLIP reductions that do not take the disk into account. We therefore use \texttt{Winnie} to initialize our disk model subtracted data into a \texttt{spaceKLIP} compatible database, which can then be used to calculate the contrast curves. Before the contrast curves can be measured, we first input a model of the TWA 7 star using the available photometry (e.g. \citealt{Hog00,Henden16,gaia18}). Next, the raw contrast curves are calculated by measuring the image noise in small annular regions, determining the contrast threshold based on the input stellar photometry and spectral type, as well as accounting for the coronagraph mask throughput and small sample statistics \citep{Mawet14}. In this step, we mask any bright companions (e.g. C1 and C2) to ensure that they do not affect the contrast at the companion's separation.

\begin{figure}[t]
    \centering
    \includegraphics[width=0.472\textwidth]{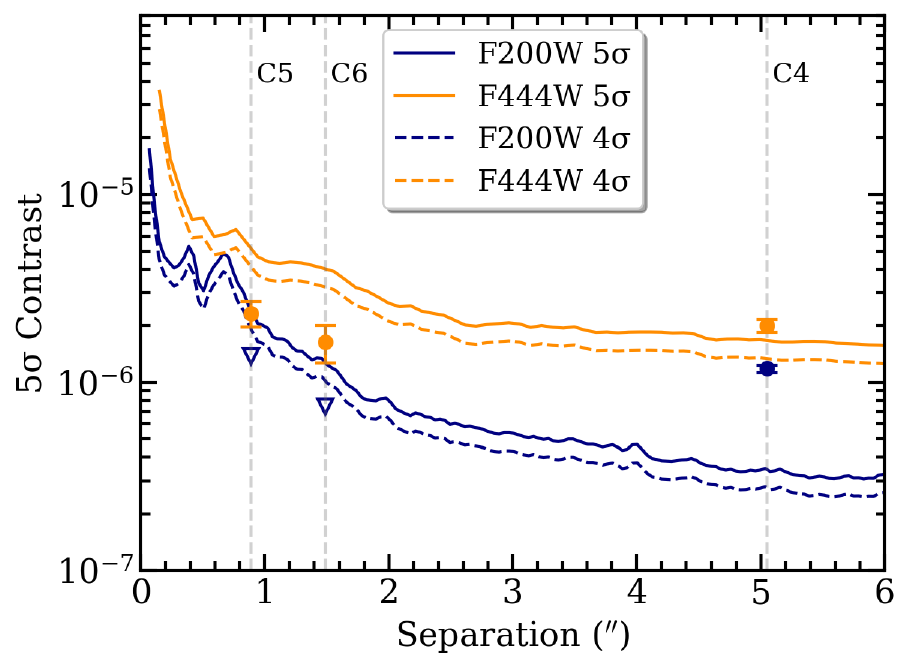}
    \caption{\label{fig:contrast_cand} 5$\sigma$ contrast curves for the F200W (navy) and F444W (orange) filters. The dashed lines represent 1$\sigma$ deviations from the $5\sigma$ contrast curves. The orange data points represent the contrasts of the candidates, C4-C6, in the F444W filter, while the navy data point represents the contrasts of the candidate, C4, in the F200W. The navy triangles represents the 3$\sigma$ upper limits placed on candidates C5 and C6 in the F200W.} 
\end{figure}

Once the raw contrasts are computed, typically the calibrated contrasts would then be measured in order to take into account other artifacts inherent to HCI (e.g. over-subtraction of companions during the PSF-subtraction process). However, \texttt{spaceKLIP} calculates the calibrated contrast curves assuming that the data were post-processed through its own KLIP algorithm, and therefore would not be able to make meaningful measurements on the flux loss through applying MCRDI with \texttt{Winnie}. In this case, instead of calculating the calibrated contrast curves, we opt to use the raw contrast curves, as they still provide improved contrasts at the location of the disk. Our 5$\sigma$ contrast curves with 1$\sigma$ deviations can be seen in Figure \ref{fig:contrast_cand}.



\subsubsection{Companion modeling} \label{sec:cand_mod}
To learn more about the three candidates that have not yet been identified as background objects (C4-C6), we first fit a model to each of the three candidates to place constraints on their ($\delta$x,$\delta$y) positions and fluxes in $\mu$Jy. Similar to Section \ref{sec:mods}, we utilize code from \texttt{Winnie} to perform forward modeling for each companion candidate (using a point-source model) alongside our best-fit disk model. In some cases it may be necessary to model both the companions and disk simultaneously, as companions close to the disk may bias the disk model. However, because we use the F200W data to create our disk model, which does not show evidence of C5 or C6 (i.e. the only two candidates that overlap with the disk), we find it sufficient to leave the parameters of our best-fit model fixed, while only the planet position and flux are allowed to vary. Given that the disk model does not take into account the asymmetries in the disk seen at F200W, we note that uncertainties in the disk model may still add additional uncertainties to the flux of the candidates, although we expect these uncertainties to be small, as the disk is far less bright and asymmetric in F444W (we explore the effect of the disk model in Section \ref{sec:disk_flux}). 

\begin{figure*}
    \centering
    \includegraphics[width=\textwidth]{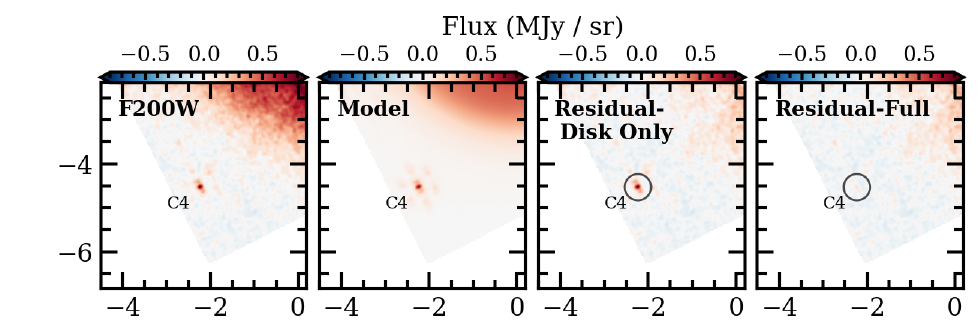}
    \includegraphics[width=\textwidth]{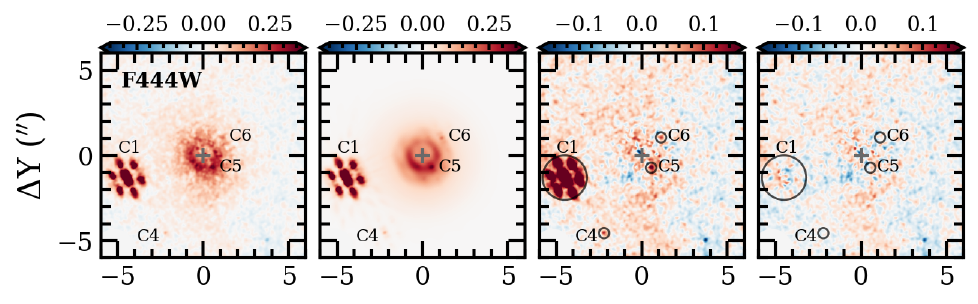}
    \includegraphics[width=\textwidth]{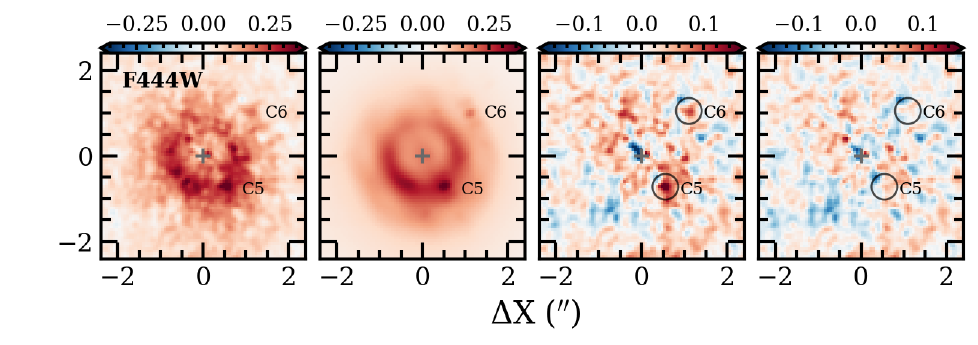}
    \caption{\label{fig:cand_mod2} \textbf{Top:} Candidate+disk models for the F200W filter. Left is our NIRCam observations, middle left is our best fit models, middle right is the residual maps with just the disk model subtracted, and right is the full residual maps with the disk and candidate models subtracted. The figure is zoomed in on candidate C4 (labeled), which is the only new companion candidate seen in the F200W. \textbf{Middle:} Companion+disk models for the F444W filter, including models for C4, C5, and C6 (labeled). \textbf{Bottom:} Same as middle panel, but zoomed in within 2.4$''$ to better view C5 and C6. The background star, C1, is also included in the forward model fitting for both filters. The circles in the residual maps indicate the location of each companion.} 
\end{figure*}

The python code, \texttt{lmfit}, is again used to minimize the residuals between the F200W/F444W data and our companion+disk model. Since we are only measuring the companion position and flux, we use the more computationally expensive ``emcee" optimization method, which utilizes Markov Chain Monte Carlo (MCMC) to calculate both the best-fit parameters and their 1$\sigma$ uncertainties. The position and fluxes for all three candidates are measured in F444W, while the position and flux for C4 are then measured in F200W. In both cases, we use visual estimates of the position and flux as initial priors and run a total of 5,000 models. Because C5 and C6 are not present in the F200W data, we only use the F444W data to estimate the position of the two candidates. Additionally, we use the F200W contrast curves to place 3$\sigma$ upper limits on the flux in $\mu$Jy for C5 and C6, although we again note that these upper limits may be biased by uncertainties in the disk model. The best-fit companion+disk models can be found in Figure \ref{fig:cand_mod2}, while the derived positions and fluxes for each candidate and their 1$\sigma$ uncertainties can be found in Table \ref{tab:candidate_mod_sum}. The 1$\sigma$ uncertainties stated for the measured F444W fluxes for C5 and C6 (which overlap with the disk) include uncertainties introduced by the disk model, which we derive in the following Section \ref{sec:disk_flux}.

Additionally, using the derived fluxes, we calculate the contrast for each candidate. This is done simply by taking the measured fluxes and uncertainties of each candidate in $\mu$Jy and dividing by the flux of the star at the F200W and F444W wavelengths in $\mu$Jy, which can be calculated in \texttt{spaceKLIP} based on the input stellar model. The resulting contrasts can be seen plotted on top of our 5$\sigma$ contrast curves in Figure \ref{fig:contrast_cand} and are added to Table \ref{tab:candidate_mod_sum} in units of apparent magnitude. C5 and C6, which are only detected at $\sim$2-3$\sigma$, sit below the 5$\sigma$ contrast curve in the F444W filter, while C4 sits above the 5$\sigma$ contrast curve in both filters.

\begin{table*}
	\centering
	\caption{\label{tab:candidate_mod_sum}Summary of the forward modeling results for candidates C4-C6. Values listed for C5 and C6 in the F200W are 3$\sigma$ upper limits, although they do not take into account uncertainties introduced from the disk model. However, these uncertainties are considered within the 1$\sigma$ uncertainties derived for the C5 and C6 fluxes in F444W.}
	\begin{tabular*}{\textwidth}{c @{\extracolsep{\fill}} ccccc}
	    \hline
	    \hline
		Name & Filter & $\delta$x ($''$) & $\delta$y ($''$) & Flux ($\mu$Jy) & Apparent Mag \\
		\hline
            \multirow{ 2}{*}{C4} & F444W & -2.23$\pm$0.19 & -4.53$\pm$0.33 & 0.92$\pm$0.07 & 20.75$\pm$0.08 \\
             & F200W & -2.24$\pm$0.02 & -4.53$\pm$0.01 & 1.51$\pm$0.07 & 21.75$\pm$0.05 \\
            \hline
            \multirow{ 2}{*}{C5} & F444W & 0.53$\pm$0.05 & -0.70$\pm$0.06 & 1.07$\pm$0.17 & 20.59$\pm$0.17 \\ 
             & F200W & ... & ... & $<$1.75 & $<$21.59 \\
            \hline
            \multirow{ 2}{*}{C6} & F444W & 1.11$\pm$0.13 & 1.00$\pm$0.09 & 0.75$\pm$0.17 & 20.98$\pm$0.25 \\
             & F200W & ... & ... & $<$0.95 & $<$22.25 \\
		\hline
		\hline
	\end{tabular*}
\end{table*}

\subsection{Disk Model Effects on Candidate Fluxes in the F444W} \label{sec:disk_flux}
We have taken two different approaches for modeling the disk and the planet candidates. Due to the number of free parameters in our disk model (24 total), we opted to use the ``Powell" optimization method, which we found to be able to derive a model that fits both the F200W and F444W data reasonably well in a time-efficient manner. To derive position and flux measurements with 1$\sigma$ uncertainties for the three planet candidates (9 parameters total), we opted to use the ``emcee" optimization method. However, by modeling the disk and planet candidates separately, this makes it difficult to estimate the uncertainty in the candidate fluxes introduced by the disk model. This is especially important for C5 and C6 which overlap the disk. 

To better quantify this uncertainty for C5 and C6, we model both the disk and the two planet candidates simultaneously in the F444W. To reduce the number of free parameters, we only vary the disk parameters most likely to differ between the F200W and F444W observations, such as the scattering phase function (i.e. $g_{1}$, $g_{2}$, and $wg_{1}$) and relative flux between the three disk components ($F$). We also allow the $PA$ for each disk component to vary several degrees to account for any uncertainty in the $PA$ measured in our initial model and because even a slight change in $PA$ may have a significant impact on the measured flux for the candidates. For this modeling run, we focus on the inner 3.5$''$ region, therefore covering the full disk plus C5 and C6. Additionally, we fix the positions for the two planet candidates to the best fit values found in Section \ref{sec:cand_mod}, so only the flux is allowed to vary. Finally, in order to minimize the computational time, we again use the ``Powell" optimization method.

The resulting best fit model varies slightly from our initial best fit model, impacting the flux of C5 and C6, in two main ways. One, our new model favors a slightly brighter outer disk component in the F444W, resulting in a slightly lower flux for C6 of 0.60 $\mu$Jy (compared to 0.75 $\mu$Jy). Second, the new model favors a slightly lower $PA$ for Ring 1 of 85$^{\circ}$, causing an increase in flux for C5 from 1.07 $\mu$Jy to 1.25 $\mu$Jy. Hence, for both C5 and C6, the change in the disk model has impacted the measured fluxes in the F444W by 15-20\%. While this is not insignificant, it is important to note that these differences are only slightly outside the 1$\sigma$ uncertainties measured for the flux in Section \ref{sec:cand_mod} (0.14 $\mu$Jy and 0.13 $\mu$Jy for C5 and C6, respectively).

Despite that the new model causes a decrease in the flux of C6 by 0.15 $\mu$Jy, there is still one more scenario to explore. This last scenario considers the fact that the disk harbors an underdense region (the ``dip" described in Section \ref{sec:ring2}) at the location of C6. This underdense region is not taken into account in our axisymmetric disk model, leading to a significant negative residual in the F200W (see Figure \ref{fig:twa7}). To measure how this dip in surface brightness affects the flux of C6 in the F444W, we create one final model. This is done by taking our initial best fit model and decreasing the relative flux parameter, $F$, for Rings 2 and 3 (the two disk components overlapping with C6) until the underdense region in the model is of a similar brightness to the F200W observations. We then use these new $F$ values to to rerun the modeling in the F444W, allowing the flux for C6 to vary. Due to the decrease in disk flux for Ring 2 and 3, the measured flux for C6 increases from 0.75 $\mu$Jy to 0.85 $\mu$Jy.

To ensure that the 1$\sigma$ uncertainties for the C5 and C6 flux include the uncertainty introduced by the disk model, we take the standard deviation of the candidate fluxes between our new models and our initial model (0.09 $\mu$Jy and 0.10 $\mu$Jy for C5 and C6, respectively) and combine them in quadrature with the 1$\sigma$ uncertainties derived in Section \ref{sec:cand_mod}. The combined uncertainties are stated in Table \ref{tab:candidate_mod_sum}.

\subsection{Planet mass estimates}

\begin{figure*}
    \centering
    \includegraphics[width=0.66\textwidth]{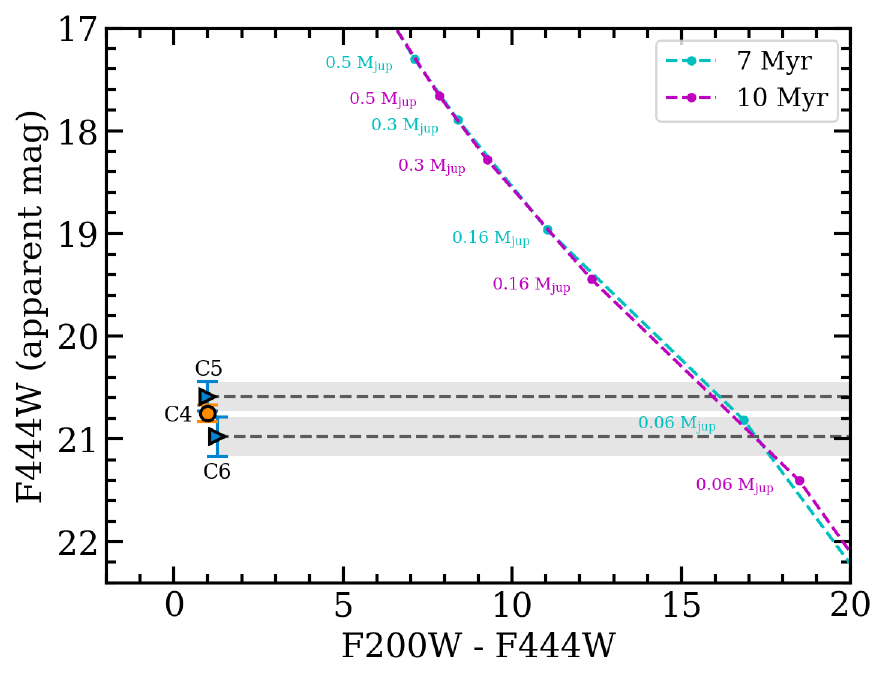}
    \includegraphics[width=0.332\textwidth]{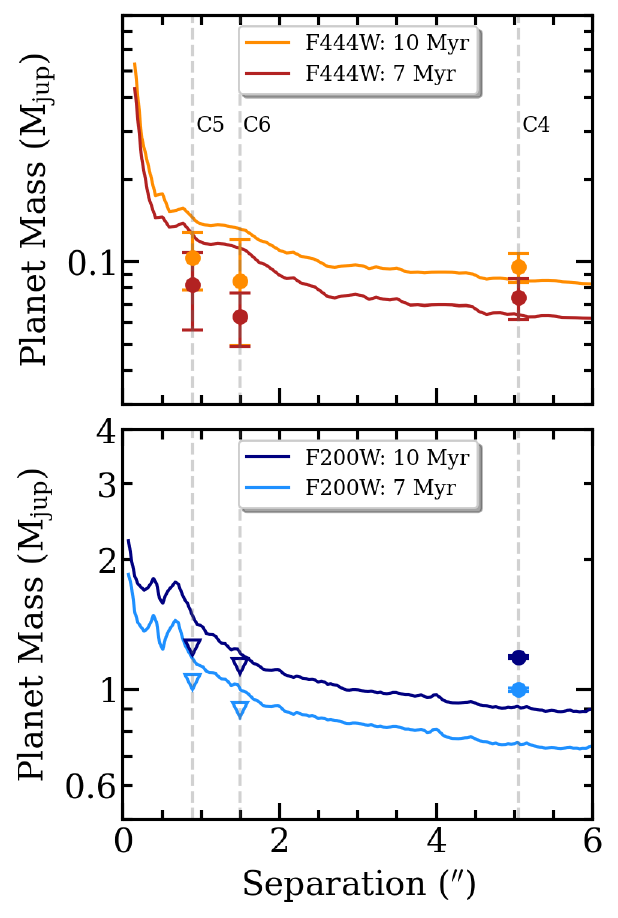}
    \caption{\label{fig:cmd} \textbf{Left:} Color-magnitude diagram with F200W-F444W vs. F444W. The orange data point represents C4, while the two blue triangles represents C5 and C6, where the color measurement is a lower limit. The cyan dashed line is the measurements derived from the \citealt{Linder19} planet evolutionary models for a system with an age of 7 Myr, while the magenta dashed line is for a system with an age of 10 Myr. The grey dashed lines and shaded regions show the parameter space possible for C5 and C6 within the 1$\sigma$ uncertainties. Using this plot, we can effectively rule out C4 being a planet companion, as it is significantly too blue in color. However, further investigation is warranted for C5 and C6 which only have lower limits for the color. \textbf{Right:} Planet mass sensitivity curves for the F444W filter (top) and the F200W filter (bottom) calculated from the 5$\sigma$ contrast curves. Error bars represent 3$\sigma$ uncertainties. For both filters the different colors represent the results derived for a system with age 7 Myr and 10 Myr.} 
\end{figure*}

With the 5$\sigma$ contrast curves and companion fluxes derived, we can use these measurements to estimate the planet mass sensitivity of our observations, as well as the potential masses of each candidate assuming they are planetary in nature. This is done by comparing the flux of the candidates to the flux predicted by planet evolutionary models. Because the estimated planet masses are likely below 1 M$_{\rm jup}$, we utilize the BEX models from \citet{Linder19}. Specifically, we use the cloud-free models with a nominal post-formation luminosity and a solar metallicity ([M/H]=0), calculated by \texttt{petitCODE}. This model is chosen as it has an extensive set of evolutionary tracks and is the standard model used for estimating the planet mass sensitivity of JWST observations (e.g. \citealt{Carter23,Franson24,Beichman25}). The measured fluxes in absolute magnitude are obtained from the planet evolutionary models for various planet masses in the F200W and F444W filters, which are then converted to apparent magnitudes based on the distance to the TWA 7 system (34.1 pc, \citealt{gaia21}). We also compare two models with ages 7 Myr and 10 Myr based on the system's estimated age ($6.4 \pm 1$ Myr, \citealt{Binks20} and 7-13 Myr, \citealt{Bell15,Luhman23}). 

To compare the estimated apparent magnitude of our companion candidates to the derived apparent magnitude of the planet evolutionary models, we create a color-magnitude diagram (CMD) using the F200W and F444W mags. The resulting CMD can be seen in the left panel of Figure \ref{fig:cmd}. Additionally, we convert our 5$\sigma$ contrast curves for both filters using the same \citep{Linder19} models, again for a system age of 7 and 10 Myr to compare. These planet mass sensitivity curves can be seen in the right panels of Figure \ref{fig:cmd}. The derived planet masses are also listed in Table \ref{tab:cand_mass}. As expected, the planet masses estimated for our candidates are below 1 Jupiter mass (with the exception of C4 in F200W), although what is surprising is that we are deriving masses as small as Neptune ($\sim$0.06-0.08 M$_{\rm jup}$). This is true even in the 10 Myr case, which predicts a higher planet mass compared to 7 Myr. If one or more of our candidates are indeed Neptune mass planets, this would make them significantly smaller than any other planet companion directly imaged to date. That being said, these estimated planet masses may be highly uncertain, as there are currently limited models in this temperature regime. These models also lack empirical validation given that there have yet to be young planets detected at these small masses with direct imaging. It is therefore possible that our potential planet candidates are cloudier or warmer than expected, meaning that the two-band photometry based on these models would likely underestimate the mass of the planet. We provide further discussion on each of the three possible planet candidates in the following section.

\begin{table}
	\centering
	\caption{\label{tab:cand_mass}Planet mass estimations for candidates C4-C6 based on clear planet evolutionary models from \citealt{Linder19}. Errorbars represent 3$\sigma$ uncertainties.}
	\begin{tabular}{cccc}
	    \hline
	    \hline
		Name & Filter & Age (Myr) & Planet Mass (M$_{\rm jup}$) \\
		\hline
            \multirow{4}{*}{C4} & F444W & 7 & 0.07$\pm$0.01 \\
             & ... & 10 & 0.10$\pm$0.01 \\
             & F200W & 7 & 1.00$\pm$0.01 \\
             & ... & 10 & 1.19$\pm$0.01 \\
             \hline
            \multirow{4}{*}{C5} & F444W & 7 & 0.08$\pm$0.03 \\
             & ... & 10 & 0.10$\pm$0.03 \\
             & F200W & 7 & $<$1.04 \\
             & ... & 10 & $<$1.25 \\
            \hline
            \multirow{4}{*}{C6}& F444W & 7 & 0.06$\pm$0.01 \\
             & ... & 10 & 0.09$\pm$0.04 \\
             & F200W & 7 & $<$0.90 \\
             & ... & 10 & $<$1.13 \\
		\hline
		\hline
	\end{tabular}
\end{table}

\section{Planet companions?} \label{sec:planets}
In Section \ref{sec:companions}, we have identified six sources between the F200W and F444W filters within $\sim$10$''$ of the TWA 7 system. Three of these sources, C1-C3, have already been identified as background objects based on prior observations, the system's proper motion, and the candidate's spatial extension. However, three faint sources, C4-C6, need further examination and possible follow-up to determine if they are background objects, planetary companions, or dust clumps a part of the disk. In this section, we take a closer look at these three potential candidates to help determine their nature. 

\subsection{C4}
C4 is the most widely-separated ($\sim$5$''$ or 170.5 au) of the three possible candidates and the only one of the three that is present in both the F200W and F444W filters, as well as detected above 5$\sigma$. It is also located near the dust clump observed in the disk, raising questions about whether the two may be connected. However, there are reasons to suspect that C4 is not consistent with a planet companion. For one, because of the strong detection in the F200W, C4 is significantly bluer than any of the planet masses predicted by the evolutionary models. This leads to two different planet mass estimates between the F444W and F200W filters of $\sim$0.1 M$_{\rm jup}$ and $\sim$1 M$_{\rm jup}$, respectively. We can therefore confidently conclude that C4 is most likely not a planetary object, but rather a faint background star or galaxy. 

\subsection{C5}
\begin{figure}[t!]
    \centering
    \includegraphics[width=0.472\textwidth]{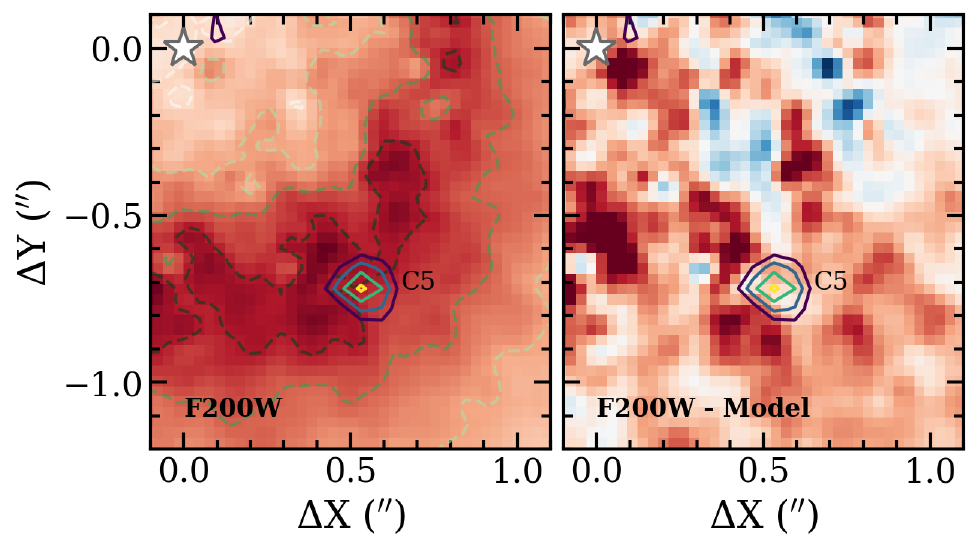}
    \caption{\label{fig:c5} Zoom in on the F200W data (left) and the model subtracted residuals in the F200W (right) around the location of C5. The green dashed contours in the left plot represent the surface brightness of the disk in F200W (between 1 and 4 MJy/sr), while the solid multicolor contours trace the flux of C5 in the F444W (between 0.13 and 0.19 MJy/sr). The white star represents the location of the TWA 7 star.} 
\end{figure}

C5 is the closest ($\sim$0.9$''$ or 30.7 au) candidate compared to C4 and C6. It is also the most complicated candidate, as it is almost completely embedded within the disk, located on the outer skirts of the innermost ring. C5 is a potentially very interesting candidate as we only have a lower limit for the color (F200W - F444W $\gtrsim$1.01), with no visible counterpart in the F200W filter, as would be expected for a planet companion. In fact, C5 is located within a heavy residual area of the F200W data, but lies directly between the residuals, as seen in Figure \ref{fig:c5}. If C5 was a dust clump within the disk, it should also be seen in F200W as we would expect the dust clump to have a similar color as the disk. Furthermore, C5 is not seen in polarized light with GPI \citep{Esposito20} or SPHERE \citep{Ren21}, which would be the case for a dust clump scenario, although it is relatively the same location as the spiral arm observed with SPHERE in 2017. It therefore seems unlikely that C5 is a dust clump, unless it has a very different composition than the rest of the disk, as was suggested for the cat's tail in the $\beta$ Pic disk \citep{Rebollido24}. However, it is also not clear whether C5 could be a companion planet. If it is indeed a planet, the magnitude of C5 would put it in the sub-Saturn range (see Table \ref{tab:cand_mass}). Given the location of C5 and the fact that it does not appear in any other observation, including MIRI, followup is necessary to determine whether or not C5 is (1) real and not an artifact, and (2) a planet or background source in the case that it is real. 

\subsection{C6}
\begin{figure}
    \centering
    \includegraphics[width=0.472\textwidth]{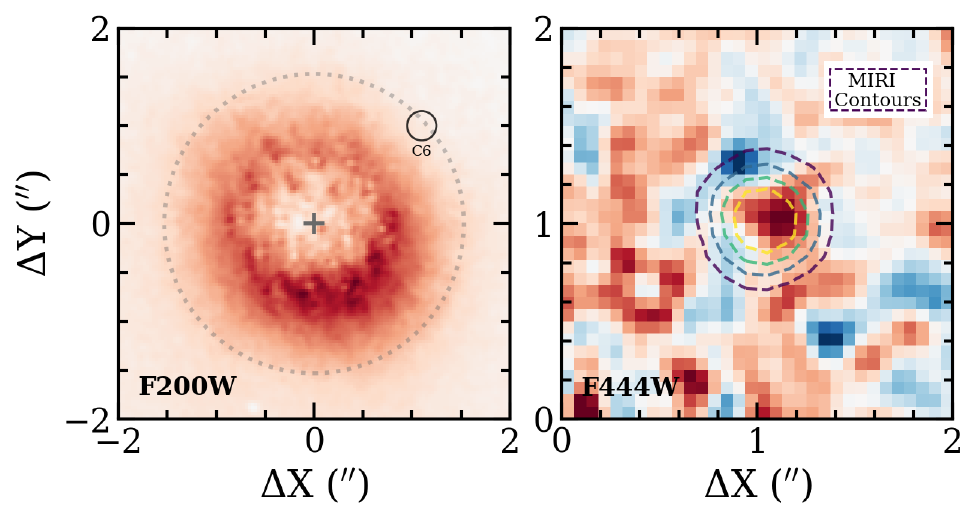}
    \caption{\label{fig:miri} \textbf{Left:} F200W observations highlighting the location of Ring 2 with the dashed grey line, as well as the location of C6 with the solid black line. \textbf{Right:} Zoom in on the candidate C6 in the F444W. The contours represent the planet candidate imaged at 11 microns with MIRI \citep{Lagrange25}.} 
\end{figure}

C6 is a particularly exciting candidate as (1) it has no counterpart in the F200W, consistent with the red colors expected for a planet companion (2) it has been imaged at a much higher SNR with MIRI and (3) if C6 is a planet, it would be directly connected to the observed disk structure. The 11.4 $\mu$m MIRI data, published recently by \citet{Lagrange25}, revealed a bright point-source like object at the same location of C6 (Figure \ref{fig:miri}). Although not entirely ruled out as a background galaxy, \citet{Lagrange25} show that the probability of this scenario, based on the flux in MIRI and upper limits from SPHERE and ALMA, is quite low ($\sim$0.3\%). What makes C6 and its MIRI counterpart even more compelling as a planet companion is its location relative to the disk. C6 is located in an underdense region of Ring 2 in the northwest quadrant, precisely where a planet has been predicted to be located by \citet{Ren21} based on Ring 2's resonant-like structure. This is demonstrated through N-body simulation work shown in \citet{Lagrange25}. Fitting atmospheric models to photometric points from MIRI and upper limits from SPHERE, \citet{Lagrange25} found that cloudy models for a 7 Myr, $\sim$0.3 M$_{\rm jup}$ (Saturn mass) planet fit these data points the best. This is roughly 5 times more massive than the planet masses derived from the \citet{Linder19} models based on the NIRCam data alone. In this case, the detection at 11 $\mu$m will be necessary to provide further constraints on the mass and atmosphere. In the following section, we combine our NIRCam results with the MIRI detection to further characterize C6 and provide evidence that it is a sub-Jupiter mass companion.

\section{Further characterization of C6} \label{sec:c6}
Through this work, we have identified three candidate companions within our NIRCam observations of the debris disk system, TWA 7. One of our candidates, C6, is detected at a much higher SNR at 11~$\mu$m with MIRI \citep{Lagrange25}. Although C6 is only detected at 2-3$\sigma$ in the F444W filter, it is located at the same position as the MIRI detection, which is also the location a planet has been predicted to reside based on the morphology of Ring 2. Given the scientific significance of the detection of a Saturn mass planet with direct imaging and within a debris disk, we focus on the C6 candidate to learn more about its nature and atmospheric properties.

\subsection{Could C6 be a background galaxy?}
So far, the photometry and upper limits obtained from SPHERE, NIRCam, and MIRI, point towards C6 being a Saturn mass companion in the TWA 7 system. However, without confirming the proper motion, there is still the possibility that this candidate could be a background object. Because the NIRCam and MIRI observations were only taken two weeks apart, this is not enough time for any significant motion of the system to occur. We must therefore utilize a different method to test the background object scenario.  

To calculate the probability of C6 being a background galaxy, \citet{Lagrange25} compared multiple SEDs of starburst and AGN galaxies of various redshifts to the measured 11 $\mu$m flux, as well as non-detections with SPHERE and ALMA. Low-intermediate redshift (0.1-1) galaxies are chosen specifically, as they are the most likely to be consistent with all three flux constraints, whereas galaxies of higher redshift would have a high likelihood of detection with ALMA. Starburst galaxies and Active Galactic Nuclei (AGN) are also specifically chosen due to the fact that C6 appears as a point-source with MIRI. Based on these constraints, \citet{Lagrange25} found that there was a $\sim$0.3\% chance of detecting such a galaxy at the location of C6. While this probability is very small, it is still non-zero and a background galaxy cannot be completely ruled out.

\begin{figure}
    \centering
    \includegraphics[width=0.472\textwidth]{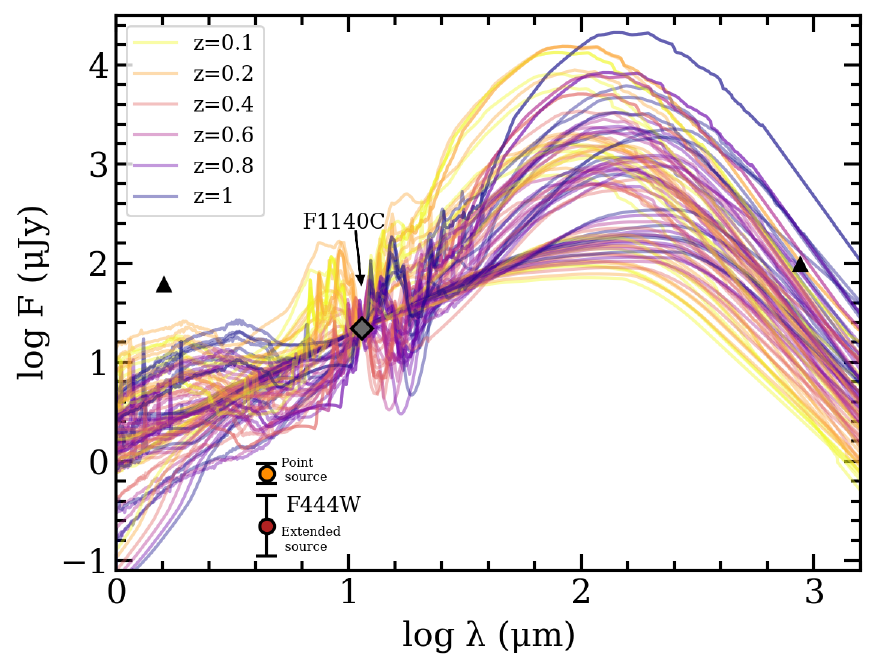}
    \caption{\label{fig:galaxy} Starburst galaxy and AGN SEDs of various redshifts consistent with the flux of C6 measured at 11 $\mu$m (grey data point) and non-detections with SPHERE and ALMA (black triangles). The red data point represents the flux of C6 measured in F444W within a 0.6$''$ aperture, while the orange data point represents the flux of C6 measured in F444W derived from the point-source modeling.} 
\end{figure}

If indeed a background galaxy, C6 should be extended in F444W. While C6 is only detected at a low SNR in F444W, it does not show significant evidence of being extended. For example, the two other background galaxies in F444W extend $\sim0.3''-0.4''$ in radius while C6  only extends $\sim 0.15''$ in radius. Additionally, when C6 is fit with a point-source model, no significant residuals remain (see bottom right panel of Figure \ref{fig:cand_mod2}). To further test the possibility of C6 being a background galaxy, we compare the F444W flux of C6 to the same galaxy SEDs used in \citet{Lagrange25}. For this, we obtain the same 14 starburst and AGN galaxy SEDs from the SWIRE Template Library\footnote{\url{https://www.iasf-milano.inaf.it/~polletta/templates/swire_templates.html}}. Similar to \citet{Lagrange25}, the flux of each galaxy SED is calibrated to the C6 flux obtained with MIRI at 11 $\mu$m. All 14 SEDs are then recalculated for various redshifts ($z = $ 0.1, 0.2, 0.4, 0.6, 0.8, 1) and are plotted in Figure \ref{fig:galaxy}. Next, we need the flux for C6 at F444W for comparison. Because C6 should be extended in F444W, we measure the photometry of C6 within an aperture with radius 0.6$''$, similar to the way the lower limit for SPHERE was measured in \citet{Lagrange25}. To calculate the uncertainty, we conduct photometry for five additional apertures with radius 0.6$''$, centered at the same radial separation as C6 at various angles. The standard deviation is then measured between the five apertures. This measurement is also shown in Figure \ref{fig:galaxy}, along with the 11 $\mu$m data point and the lower limits placed by SPHERE and ALMA. Additionally, for comparison, we plot the flux measured for C6 from the companion modeling.

Figure \ref{fig:galaxy} shows that the estimated flux of C6 in the F444W is significantly below all 14 galaxy SEDs at 4-5 $\mu$m, regardless of redshift. In the case where we measure the flux of C6 as an extended object (within a 0.6$''$ aperture), we get an even lower estimate than when measuring the flux as a point-source. This is because there are several negative residuals within 0.6$''$ of C6. While this may be mitigated by correcting the disk model, the flux would still need to increase by $\gtrsim$6.7 times to be consistent with any of the galaxy SEDs. This suggests that if C6 is indeed a background galaxy, it should have been detected at a higher SNR in F444W. Thus, our NIRCam observations provide additional evidence that C6 is more likely a planetary companion and not a background galaxy.

\subsection{Atmospheric modeling}

\begin{figure*}
    \centering
    \includegraphics[width=\textwidth]{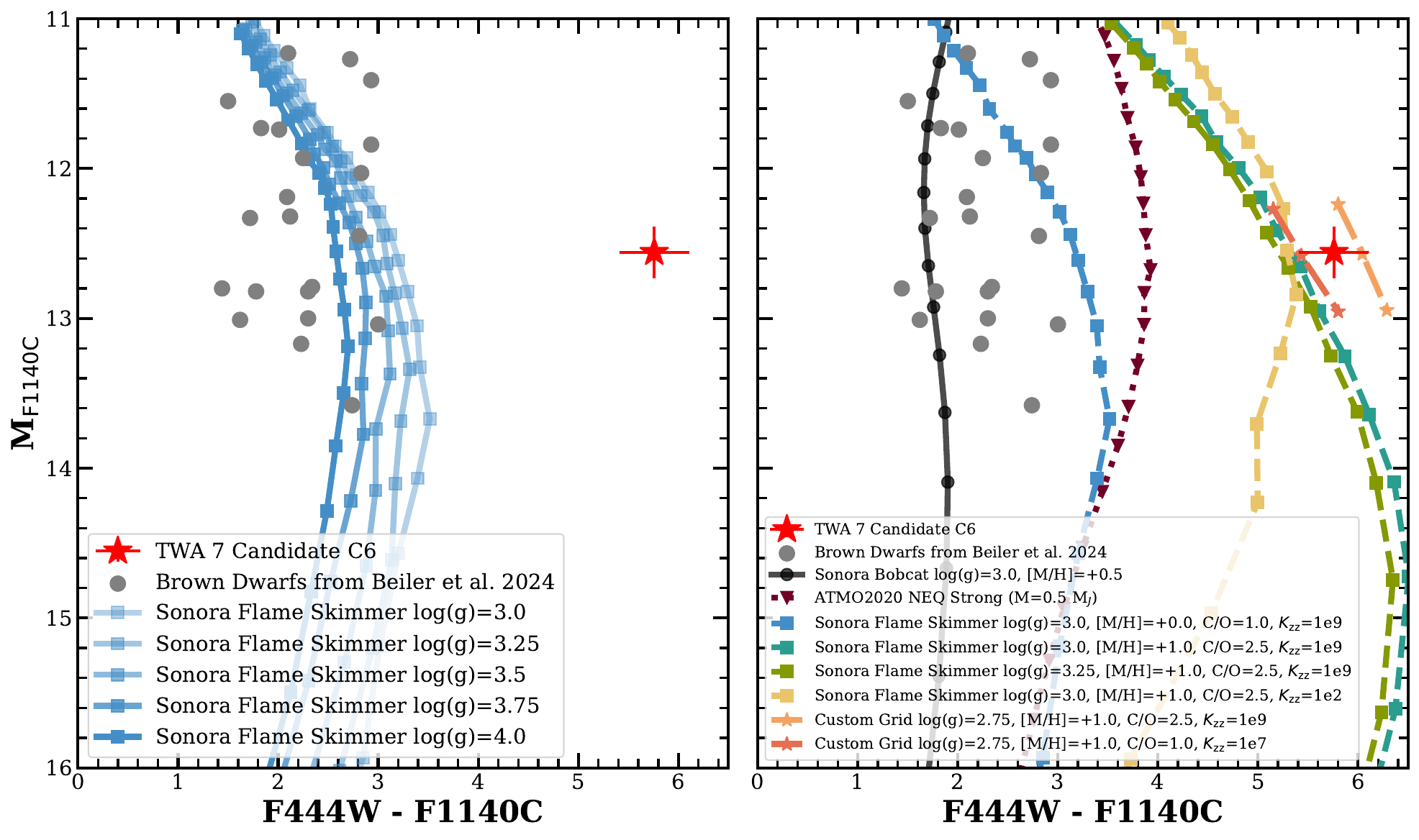}
    \caption{\label{fig:cmd2} Color-magnitude diagrams illustrating the extreme redness of the C6 companion (red). Gray points show JWST synthetic photometry of brown dwarfs from \citet{Beiler24b} (gray). Left: Selection of Sonora Flame Skimmer models with [M/H] = +0.0, C/O = 1.0, and $K_{\rm zz}$ = 1e9. Right: Atmospheric model tracks from Sonora Bobcat (solid circles) and ATMO2020 (dotted triangles) are compared to new Sonora Flame Skimmer models with enhanced metallicity (dashed lines). Custom models generated for this study are shown with dash-dotted lines.} 
\end{figure*}

Considering the evidence that C6 is planetary in nature, we compare our results against the atmospheric properties found in \citet{Lagrange25}. For this, we first examine how well existing grids of atmospheric models can match the photometric detections from both MIRI and NIRCam. We present C6 on another CMD (Figure \ref{fig:cmd2}), plotting its Vega magnitude in F1140C against the color F444W $-$ F1140C. We include synthetic JWST photometry for 22 of the 23 late-T and Y dwarfs compiled by \citet{Beiler24b}; WISEA J2018-74 is excluded since synthetic F444W photometry was not available. Compared to this brown dwarf sample, C6 stands out as significantly redder, emphasizing its unique position within the substellar population. We also overlay model tracks derived from synthetic photometry from three grids: Sonora Bobcat \citep{Marley2021}, ATMO2020 \citep{Phillips2020}, and the upcoming Sonora Flame Skimmer (Mang et al., in prep). The Flame Skimmer models extend the Sonora Elf Owl grid \citep{Mukherjee2024EO} to colder temperatures and lower surface gravities. Key updates include the incorporation of rainout chemistry for condensates such as H$_2$O—even in cloud-free atmospheres—following the treatment in Sonora Bobcat and improvement to the CO$_2$ abundance based on the chemical kinetics of \citet{Zahnle2014}, which addresses the underestimation identified in the Elf Owl models by \citet{Beiler2024}.

\begin{figure}
    \centering
    \includegraphics[width=\columnwidth]{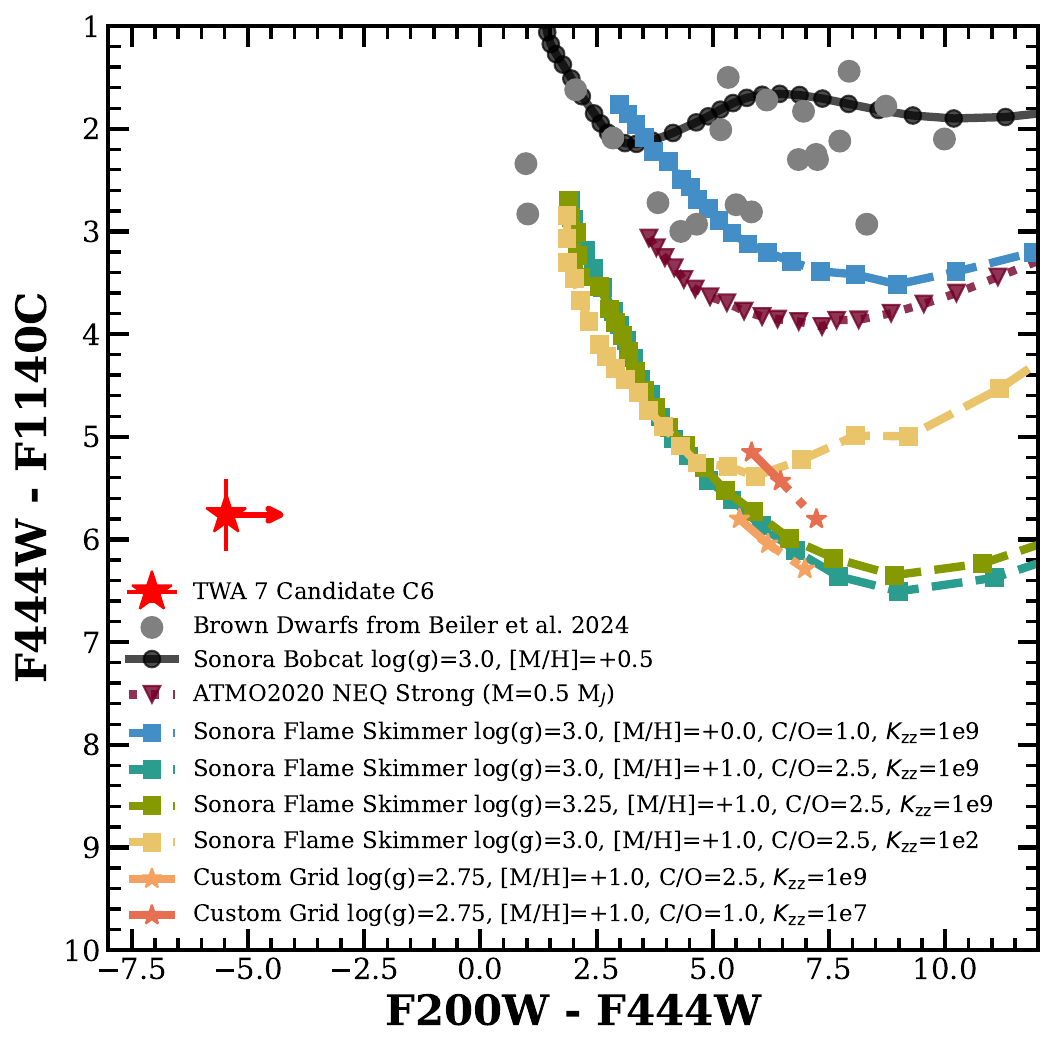}
    \caption{\label{fig:colorcolor} Color–color diagram showing the C6 companion (red) alongside a sample of brown dwarfs from \citet{Beiler24b} (gray). The F200W band for C6 represents a 3$\sigma$ upper limit. Atmospheric model tracks from Figure \ref{fig:cmd2} are overlaid, highlighting the distinct properties of C6 compared to the brown dwarf population.} 
\end{figure}

The Sonora Bobcat models, which are cloud-free and assume chemical equilibrium, are too bright to match the faint F444W flux of C6. Higher-than-solar metallicity and disequilibrium chemistry decrease the F444W flux by increasing the abundances of CO$_2$ and CO, which absorb within that bandpass. The non-equilibrium ATMO2020 models still fail to reproduce the observed faintness of C6. The Sonora Flame Skimmer models, which include vigorous vertical mixing and enhanced metallicity (10$\times$ solar), more closely match C6’s observed color, especially when extended to lower gravities in our custom grid. Looking at a color-color diagram using the F200W upper limit (Figure \ref{fig:colorcolor}), C6 is consistent with the broad range of red colors spanned by our models.

Since no current models adequately match the observed faintness of C6 in F444W, we generated a grid of custom atmospheric models to better explore this parameter space. We used \texttt{PICASO} \citep{Batalha2019, Mukherjee2023}, an open-source Python package that computes one-dimensional pressure–temperature (P–T) profiles in radiative–convective equilibrium to generate a small set of models. To include clouds, we use the \texttt{EddySed} model described by \citet{Ackerman2001}, a method widely used in other studies \citep{marley2010, skemer2016, Morley2015, Rajan2017} and utilized in model grids such as those in \citet{Saumon2008} and the Sonora Diamondback series \citep{Morley2024}. Within \texttt{PICASO}, we implement an updated Python-based version of the \cite{Ackerman2001} model, called \texttt{Virga} \citep{virga}. 

\begin{figure*}
    \centering
    \includegraphics[width=\textwidth]{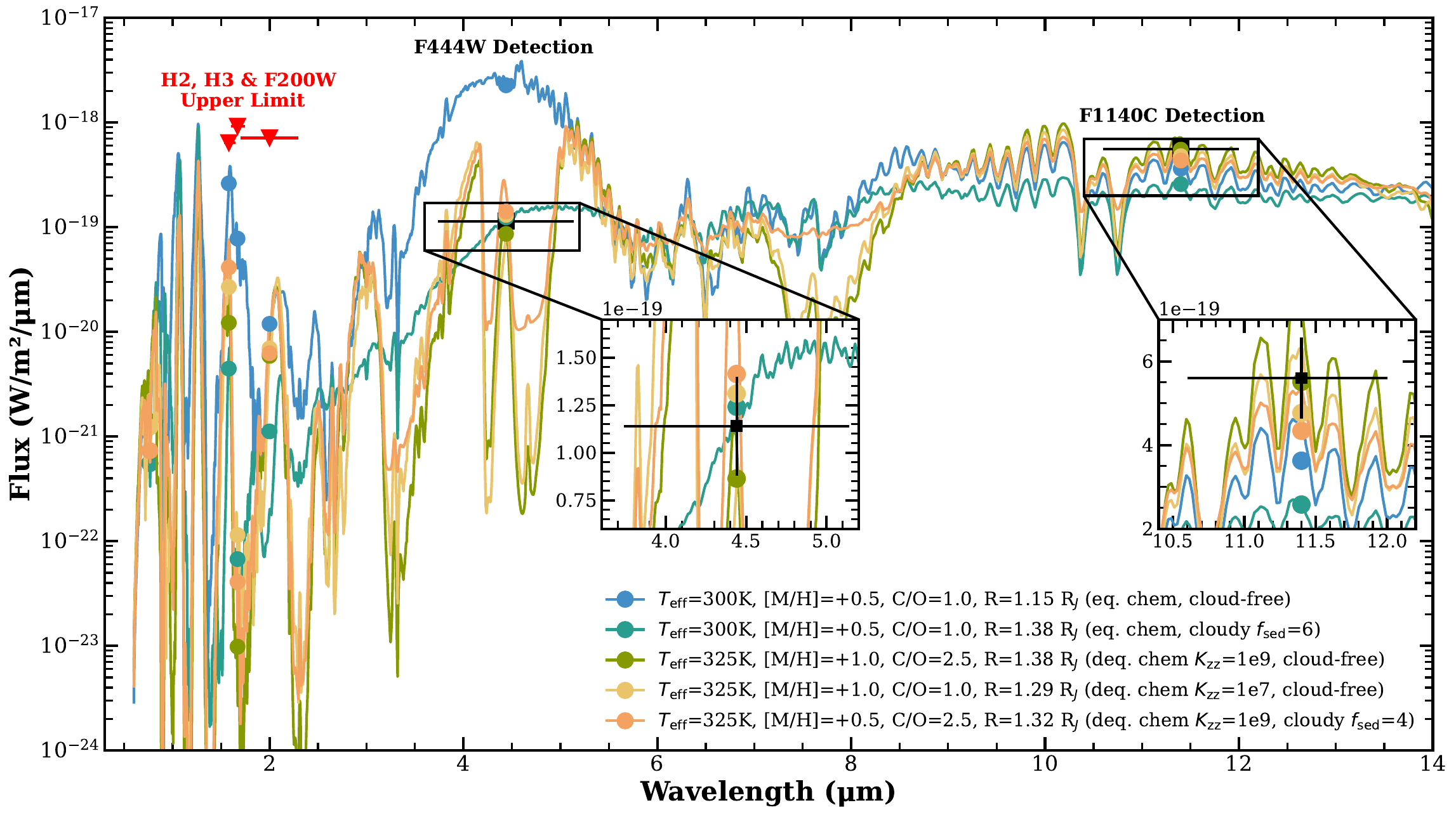}
    \caption{\label{fig:c6_mod} Custom atmospheric models generated with \texttt{PICASO} to fit the JWST NIRCam F444W and MIRI F1140C photometric detections of C6, along with the 3$\sigma$ upper limit at F200W and the 5$\sigma$ upper limits in the H2 and H3 filters from SPHERE. All models assume a surface gravity of log(g) = 2.75 (cgs), with the object's mass ($\sim$0.3 M$_{\rm jup}$) varying slightly to maintain this surface gravity and the corresponding radius. The best-fitting models within the 1$\sigma$ uncertainty correspond to 325 K planets with 10$\times$ solar metallicity. The C/O ratio and radius can vary to match the observed photometry. The two insets provide a detailed look at each photometric detection.} 
\end{figure*}

Our small grid of models spans effective temperatures of $T_{\rm eff}$ = [300, 325, 350 K], surface gravity of log(g) = 2.75 (cgs), eddy diffusion parameter of $K_{\rm zz}$ = [10$^2$, 10$^7$, 10$^9$ cm$^2$ s$^{-1}$], a metallicity of [M/H] = [+0.0, +0.5, +1.0], and C/O = [1.0, 2.5] (relative to the sun). Cloudy models have an $f_{\rm sed}$ = [2, 4, 6, 8], with H$_2$O as the only condensing species. The parameter space for this custom model grid was selected based on initial comparisons of C6 with existing grids of atmospheric models (Figure \ref{fig:cmd2}). From the color–magnitude diagram, the parameters that most strongly influence the color are metallicity, vertical mixing strength ($K_{\rm zz}$), and the C/O ratio. Therefore, we fixed the effective temperature to the 1$\sigma$ range reported in \citet{Lagrange25} and adopted a surface gravity based on their best-fit mass and radius estimates while exploring a range of metallicity, $K_{\rm zz}$, and C/O ratios. We compute moderate-resolution thermal emission spectra with \texttt{PICASO}. These spectra span 0.6 to 28 $\upmu$m at a resolving power of 5000. Finally, opacities are taken from \citet{Freedman2008}, with updates from \citet{Freedman2014}, and we incorporate the revised opacity tables from \citet{Morley2024} and \citet{Mukherjee2024EO}.

The best-fit atmospheric models are shown in Figure \ref{fig:c6_mod}. During the fitting process, we explored a range of planet radii within the 1$\sigma$ limits from \citet{Lagrange25} while maintaining a planet mass of $\sim$0.3 M$_{\rm jup}$. The planet mass, also estimated from \citet{Lagrange25}, is varied slightly to maintain the assumed surface gravity of log(g) = 2.75 (cgs) and the corresponding radius. We find that no chemical equilibrium model can fully reproduce the observed brightness in both the NIRCam F444W and the MIRI F1140C band. Instead, there are two models that best match the F444W and F1140C photometric points within their 1$\sigma$ uncertainties (the yellow and green lines in Figure \ref{fig:c6_mod}). In both cases, C6 could be a cloud-free planet in chemical disequilibrium with an effective temperature of 325 K and 10× solar metallicity. One model assumes a C/O ratio of 1.0, a radius of 1.29 $R_{\rm J}$, and $K_{\rm zz}$ = 10$^7$ cm$^2$ s$^{-1}$; the other assumes a C/O ratio of 2.5, a radius of 1.38 $R_{\rm J}$, and $K_{\rm zz}$ = 10$^9$ cm$^2$ s$^{-1}$.

There are several conclusions that can be drawn from our best-fit atmospheric model. For one, despite the lower mass derived from \citet{Linder19} evolutionary models, when taking into account the MIRI detection, both the F444W detection of C6 and the F200W upper limits are consistent with a Saturn-mass planet of temperature 300-325 K and a higher than solar metallicity as derived in \citet{Lagrange25}. However, there are some key differences between our best-fit models. For example, unlike the \citet{Lagrange25} models, our two best-fit models are cloud-free, although the cloudy chemical equilibrium model (orange line in Figure \ref{fig:c6_mod}) is consistent with both the NIRCam and MIRI data points within 2$\sigma$. Additionally, our models require chemical disequilibrium to better fit the data. This result highlights the complex relationship between factors shaping the atmosphere of a cool Saturn-mass companion, where clouds, chemical (dis)equilibrium, and metallicity all contribute in ways that cannot be fully disentangled from one single photometric detection alone. In order to better model the atmosphere, additional photometric points and future observations are required; however, for now, our models demonstrate that C6 is consistent with a planetary nature. 

\section{Conclusions}
In this work, we presented new NIRCam observations of the debris disk system, TWA 7, in the F200W and F444W filters, as part of the sub-Jupiter imaging survey, GO 4050. The data were reduced with \texttt{spaceKLIP} and \texttt{Winnie} and then forward modeled in order to model and subtract the disk from the data. Based on our data and analysis, we find the following results:

\begin{itemize}
    \item The debris disk is detected in both filters, with the disk being brighter in the F200W filter. We find that the disk harbors most of the substructures seen in previous observations, including the three separate disk components, the upper ``spiral arm" (S2) and the southern dust clump. However, we do not detect the same S1 spiral arm observed with SPHERE in 2017. Additionally, the southern dust clump appears to be connected to the inner disk with a possible extension north of the dust clump.
    \item We measure the azimuthal surface brightness of Ring 2 and compare the location of the underdense region in the NW region to observations with SPHERE and STIS. Contrary to \citet{Ren21}, we find that the underdense region of Ring 2 is more consistent with either no change or a very small change of $\sim$1-2$^{\circ}$ clockwise per year or less. This level of rotation of Ring 2 is more consistent with the period of a sub-Jupiter mass planet at a separation of $1\farcs53$ ($\sim0\fdg87$ per year for a Saturn mass planet).
    \item After subtracting the disk models we identified six sources, labeled from C1-C6. C1 is a known background star, while C2 and C3 are known background galaxies. The last three sources, C4-C6, required further investigation to discriminate between a background object, dust clump, and planet candidate. 
    \item By combining both disk and point-source forward modeling, we are able to estimate the flux and position of the three potential planet candidates in F444W. Furthermore, we measure the flux of C4 and place 3$\sigma$ upper limits for C5 and C6 in F200W. Based on the flux measurements, C4 is likely a background object, as it is too blue to be a planet. C5 is consistent with a $\sim$0.1 M$_{\rm jup}$ planet, assuming cloud-free planet evolutionary models; however, further observations are required to determine the nature of C5 and whether it is real or an artifact.
    \item C6 is the most promising of the candidates, as it is located at the same position as a potential planet companion detected with MIRI at 11 $\mu$m \citep{Lagrange25}. Additionally, this companion is at the same location where a sub-Jupiter mass planet is predicted to reside based on the morphology of Ring 2 \citep{Ren21}. Based on planet evolutionary models from \citet{Linder19}, NIRCam flux measurements predict a planet of Neptune mass for C6. Although, when including measurements from MIRI, we find that the best-fit atmospheric model is consistent with a Saturn mass planet of temperature 300-325 K as found in \citet{Lagrange25}. However, we find that a cloud-free model with a chemical disequilibrium and higher metallicity best fit the data, demonstrating the importance of additional atmospheric parameters for cool planet companions beyond just clouds. Finally, if C6 was a background galaxy, we would have expected to robustly detect it F444W; hence our NIRCam data further support the interpretation of this candidate as a bonafide exoplanet companion. 
\end{itemize}

In summary, TWA 7 is a truly fascinating system, with both a face-on/complex debris disk and a planet candidate directly interacting with the disk. Although follow-up observations are required to confirm the candidate's proper motion and for further characterization, our NIRCam data strongly support the interpretation of C6 as a Saturn-mass planet, thus making it the lowest mass planet directly imaged to date. Additionally, these results help further confirm the connection between planets and their debris disks and make TWA 7 the perfect system for studying planet-disk interactions and exoplanet evolution.

\begin{acknowledgements}
The authors wish to thank the anonymous referee for helpful suggestions that improved this manuscript. This work is based on observations with the NASA/ESA/CSA JWST, obtained at the Space Telescope Science Institute, which is operated by AURA, Inc., under NASA contract NAS 5-03127. These observations are associated with the JWST program 4050 (PI: A.Carter). The JWST data presented in this article were obtained from the Mikulski Archive for Space Telescopes (MAST) at the Space Telescope Science Institute. The specific observations analyzed can be accessed via \dataset[doi: 10.17909/19et-s876]{http://dx.doi.org/10.17909/19et-s876}. Support for program 4050 was provided by NASA through a grant from the Space Telescope Science Institute, which is operated by the Association of Universities for Research in Astronomy, Inc., under NASA contract NAS 5-03127. This work benefited from the 2024 Exoplanet Summer Program in the Other Worlds Laboratory (OWL) at the University of California, Santa Cruz, a program funded by the Heising-Simons Foundation. This material is based upon work supported by the National Science Foundation Graduate Research Fellowship under Grant No.~2139433. This project has received funding from the European Research Council (ERC) under the European Union's Horizon 2020 research and innovation program (COBREX; grant agreement \#885593). C.V.M and J.M. acknowledge the support of STScI grant JWST-AR-01977.004-A, JWST-GO-02327.010-A, JWST-AR-03245.004-A, JWST-GO-03337.004-A, and JWST-GO-04050.009-A. C.V.M acknowledges support of NASA XRP grant 80NSSC24K0958. J.M. acknowledges support from the National Science Foundation Graduate Research Fellowship Program under Grant No. DGE 2137420. 

These JWST observations were supported with high resolution speckle imaging obtained via the community program for High-Resolution Imaging, PI: Steve B. Howell (NASA/Ames). This program is supported by the NASA Exoplanet Program Office. We are also grateful to Christian Ginski, Colin Littlefield, Catherine Clark, and Elise Furlan for their support with ground-based binary vetting efforts. 
\end{acknowledgements}

\facilities{JWST}

\software{spaceKLIP (\citealt{Kammerer22,Carter23}, \url{https://spaceklip.readthedocs.io/en/latest/index.html}),
Winnie (\citealt{Lawson22}, \url{https://github.com/kdlawson/Winnie}),
lmfit (\citealt{lmfit25}, \url{https://lmfit.github.io/lmfit-py/}),
matplotlib (\citealt{Hunter07}), 
iPython (\citealt{Perez07}), 
Astropy (\citealt{astropy18}),
NumPy (\citealt{Harris20}; \url{https://numpy.org}),
SciPy (\citealt{Virtanen20}; \url{http://www.scipy.org/})}

\appendix

\begin{table*}
	\centering
	\caption{\label{tab:psf_data_sum}Summary of the PSF data used in this paper. Similar to Table \ref{tab:data_sum}, $N_{ints}$ is the total number of integrations, $N_{groups}$ is the number of groups per integration, and $N_{frames}$ is the number of frames per group. Additionally, t$_{int}$ is the effective integration time in seconds and t$_{exp}$ is the total effective exposure time in seconds.}
	\begin{tabular*}{\textwidth}{c @{\extracolsep{\fill}} ccccc}
	    \hline
	    \hline
		Name & Date & Readout Pattern & $N_{ints}$/$N_{groups}$/$N_{frames}$ & t$_{int}$ (s) & t$_{exp}$ (s) \\
		\hline
            IRAS 13041-5653 & 2024 Jun 6 & RAPID & 2/5/1 & 5 & 10 \\
            HD 89063 & 2024 Jun 14 & MEDIUM8 & 2/6/8 & 62 & 124 \\
            Y Crt & 2024 Jun 15 & SHALLOW4 & 2/4/4 & 20 & 40 \\
            IRAS 09432-4847 & 2024 Jun 15 & DEEP8 & 3/7/8 & 137 & 411 \\
            i Vel & 2024 Jun 15 & DEEP8 & 8/4/8 & 73 & 582 \\
		  CD-23 9765 & 2024 Jun 15-16 & DEEP8 & 2/12/8 & 244 & 488  \\
            HD 101581 & 2024 Jun 16 & DEEP8 & 2/6/8 & 115 & 230 \\
		\hline
		\hline
	\end{tabular*}
\end{table*}

\section{PSF Library Observations}
In Section \ref{sec:obs}, we describe the PSF library used in the post-processing of the TWA 7 datasets. Table \ref{tab:psf_data_sum} lists the summary of the seven PSF observations. All observations were taken between June 6 2024 UT and June 16 2024 UT.

\clearpage


\bibliography{sample631.bbl}{}
\bibliographystyle{aasjournal}



\end{document}